\documentclass[english,aip,letter,superscriptaddress,floatfix,jcp,11pt,nofootinbib]{revtex4}
\usepackage[T1]{fontenc}
\usepackage[latin9]{inputenc}
\usepackage{ae,aecompl}
\usepackage{color}
\usepackage{verbatim}
\usepackage{float}
\usepackage{calc}
\usepackage{mathtools}
\usepackage{amsmath}
\usepackage{amssymb}
\usepackage{graphicx}
\usepackage{esint}
\usepackage{stmaryrd}
\usepackage{footmisc}
\usepackage{yfonts}
\usepackage{dsfont}
\usepackage{subfig}
\makeatletter

\floatstyle{ruled}
\newfloat{algorithm}{tbp}{loa}
\providecommand{\algorithmname}{Algorithm}
\floatname{algorithm}{\protect\algorithmname}

\@ifundefined{textcolor}{}
{%
 \definecolor{BLACK}{gray}{0}
 \definecolor{WHITE}{gray}{1}
 \definecolor{RED}{rgb}{1,0,0}
 \definecolor{GREEN}{rgb}{0,1,0}
 \definecolor{BLUE}{rgb}{0,0,1}
 \definecolor{CYAN}{cmyk}{1,0,0,0}
 \definecolor{MAGENTA}{cmyk}{0,1,0,0}
 \definecolor{YELLOW}{cmyk}{0,0,1,0}
}

\usepackage{ae,aecompl}

\usepackage{amsthm,mathtools}
\theoremstyle{plain}

\numberwithin{thm}{subsection}

\numberwithin{lem}{subsection}

\usepackage{algorithm}
\usepackage[noend]{algpseudocode}

\newcommand{\deleted}[1]{}

\makeatother

\usepackage{babel}
\begin{document}

\title{Brownian Dynamics of Fully Confined Suspensions of Rigid Particles\\ Without Green's Functions}

\author{Brennan Sprinkle}
\affiliation{Engineering Science and Applied Math, Northwestern University, Evanston, IL 60208}
\affiliation{Courant Institute of Mathematical Sciences, New York University,
New York, NY 10012}

\author{Aleksandar Donev}
\email{donev@courant.nyu.edu}
\affiliation{Courant Institute of Mathematical Sciences, New York University, New York, NY 10012}

\author{Amneet Pal Singh Bhalla}
\affiliation{Department of Mechanical Engineering, San Diego State University, San Diego, CA 92182}

\author{Neelesh Patankar}
\email{n-patankar@northwestern.edu}
\affiliation{McCormick School of Engineering, Northwestern University, Evanston,
IL 60208}

\begin{abstract}
We introduce a Rigid-Body Fluctuating Immersed Boundary (RB-FIB) method to perform large-scale Brownian dynamics simulations
of suspensions of rigid particles in fully confined domains, without any need to explicitly construct Green's functions or mobility operators.
In the RB-FIB approach, discretized fluctuating Stokes equations 
are solved with prescribed boundary conditions in conjunction with a rigid-body immersed boundary method to 
discretize arbitrarily-shaped colloidal particles with no-slip or active-slip prescribed on their surface.
We design a specialized Split--Euler--Maruyama temporal integrator that uses a combination of random finite differences to capture the stochastic drift appearing in the overdamped Langevin equation.
The RB-FIB method presented in this work only solves mobility problems in each time step using a preconditioned iterative solver,
and has a computational complexity that scales linearly in the number of particles and fluid grid cells.
We demonstrate that the RB-FIB method correctly reproduces the Gibbs-Boltzmann equilibrium distribution, and 
use the method to examine the time correlation functions for two spheres tightly confined in a cuboid.
We model a  quasi--two-dimensional colloidal crystal confined in a narrow microchannel and hydrodynamically driven across a commensurate periodic substrate potential mimicking the effect of a corrugated wall. We observe partial and full depinning of the colloidal monolayer from the substrate potential above a certain wall speed, consistent with a transition from static to kinetic friction through propagating kink solitons. Unexpectedly, we find that particles nearest the boundaries of the domain are the first to be displaced, followed by particles in the middle of the domain.
\end{abstract}

\maketitle
\global\long\def\V#1{\boldsymbol{#1}}
\global\long\def\M#1{\boldsymbol{#1}}
\global\long\def\Set#1{\mathbb{#1}}

\global\long\def\D#1{\Delta#1}
\global\long\def\d#1{\delta#1}

\global\long\def\norm#1{\left\Vert #1\right\Vert }
\global\long\def\abs#1{\left|#1\right|}

\global\long\def\grad{\M{\nabla}}
\global\long\def\avv#1{\langle#1\rangle}
\global\long\def\av#1{\left\langle #1\right\rangle }

\global\long\def\P{\mathcal{P}}
\global\long\def\msd{\V{\text{MSD}}}

\global\long\def\Dm{\M{D}}
\global\long\def\Gm{\M{G}}
\global\long\def\DTm{\pmb{\mathds{D}}}
\global\long\def\GTm{\pmb{\mathds{G}}}
\global\long\def\Lm{\pmb{\mathbb{L}}}

\global\long\def\ki{k}
\global\long\def\wi{\omega}

\global\long\def\slip{\breve{\V u}}

\global\long\def\bu{\V u}
 \global\long\def\bv{\V v}
 \global\long\def\br{\V r}

\global\long\def\sM#1{\M{\mathcal{#1}}}
\global\long\def\Mob{\sM M}
\global\long\def\J{\sM J}
\global\long\def\S{\sM S}
\global\long\def\L{\sM L}

\global\long\def\N{\sM N}
\global\long\def\K{\sM K}
\global\long\def\slipN{\breve{\N}}

\global\long\def\aN{\widetilde{\N}}
\global\long\def\aK{\widetilde{\K}}
\global\long\def\aMob{\widetilde{\Mob}}

\global\long\def\epsN{\overline{\N}}

\global\long\def\slipW{\breve{\V W}}
\global\long\def\rot{\M{\Psi}}
\global\long\def\Rot{\M{\Xi}}

\section{Introduction}

Suspensions of micron sized rigid colloidal particles can be investigated by an ever expanding set of experimental tools such as optical tweezers and externally applied gravitational, magnetic and electric fields. Passive and active colloidal suspensions studied in the lab or in nature are almost always confined to be near one or more physical boundaries such as microscope slips \cite{ConfinedSphere_Sedimented,BoomerangDiffusion}. The confinement geometry strongly influences the underlying hydrodynamics of micro-particle suspensions due to the long-ranged nature of hydrodynamic interactions in the steady Stokes regime \cite{AutophoreticSpheres_Adhikari}, and boundaries may also play a key role in propulsion mechanisms in active suspensions \cite{ActiveDimers_EHD}. In recent years, precise and detailed measurements have been performed on quasi--two-dimensional (quasi--2D) colloidal crystal monolayers in order to interrogate kinetic friction at the microscopic scale \cite{Bohlein2011,ColloidSheetFriction}, but numerical simulations of this kind of system with proper accounting for hydrodynamic interactions are lacking. In this paper we introduce a numerical method that can simulate Brownian suspensions of rigid colloidal particles of complex shape in fully confined rectangular domains, with arbitrary combinations of periodic, no-slip, or free-slip boundary conditions along different dimensions.


Designing scalable simulation techniques for Brownian suspensions of many passive or active non-spherical colloids in confined domains is still an outstanding challenge in the field. While a number of existing methods can efficiently handle triply-periodic domains using Ewald techniques \cite{BrownianDynamics_OrderNlogN,StokesianDynamics_Brownian,FluctuatingFCM_DC,SpectralSD}, many experiments are carried out in some form of tight confinement, such as the dynamics of particles pressed between two microscope slides \cite{ColloidsInTightSlit,BoomerangDiffusion} or flowing through a microchannel. Stokesian Dynamics (SD) has become an industry standard in chemical engineering circles for Brownian suspensions \cite{BrownianDynamics_OrderNlogN,StokesianDynamics_Brownian,SD_SpectralEwald,SpectralSD}, and the method has been adapted to non-spherical rigid particles \cite{RigidBody_SD,StokesianDynamics_Rigid,StokesianDynamics_Confined}, as well as particles in partial or full confinement \cite{StokesianDynamics_Confined,StokesianDynamics_Wall,StokesianDynamics_Slit}. However, traditional SD and recent closely-related boundary-integral  \cite{AutophoreticSpheres_Adhikari,BoundaryIntegralWall_Adhikari} `implicit-fluid' methods remain limited to spherical particles in specific geometries, for which a (grand) mobility tensor can be constructed explicitly. Furthermore, existing methods only have computational complexity that scales linearly in the number of particles for periodic boundary conditions, for which Fast Fourier Transforms (FFTs) accelerate the many-body computations \cite{BrownianDynamics_OrderNlogN,StokesianDynamics_Brownian,FluctuatingFCM_DC,SpectralSD}.
Fluctuating Lattice Boltzmann simulations have also been used for suspensions \cite{LB_SoftMatter_Review}. These techniques, however, require introducing artificial fluid compressibility and fluid inertia, which imposes a severe restriction on the time step size in order to achieve a physically-realistic Schmidt number.

The need to explicitly construct and compute Green's functions for nontrivial boundary conditions can be avoided by using a grid-based solver for the fluctuating Stokes equations in order to compute the action of the Green's function \emph{and} generate Brownian forces.
The authors of \cite{SELM_FEM} presented a Finite Element Method (FEM) capable of simulating Brownian suspensions in very general, confined domains. This method however, was limited to minimally resolved (point-like) spherical particles and did not correctly account for the stochastic drift term that appears in the overdamped Langevin equations when the mobility is configuration dependent. The authors of \cite{DECORATO_FEM} proposed a FEM scheme with body-fitted grids that is capable of simulating Brownian particles with arbitrary shape in confined domains. However, the method requires complex remeshing every time step and the temporal integrator used in this work requires solving expensive resistance problems. Because of this, the body-fitted FEM approach does not scale well in the number of particles and is in practice limited to one or a few individual particles.

The Rigid-Body Fluctuating Immersed Boundary Method (RB-FIB) method presented in this work is, to our knowledge, the first method that can simulate Brownian suspensions of rigid particles with arbitrary shape in fully confined domains with controllable accuracy and in computational time that scales linearly in the number of particles (at finite packing fractions). The method is built on contributions from a number of past works by some of us. In \cite{BrownianBlobs}, the authors presented a Fluctuating Immersed Boundary (FIB) method which could simulate fluctuating suspensions of minimally resolved spheres (or blobs) in general physical domains. In \cite{RigidMultiblobs}, the authors presented a rigid multiblob method to simulate \emph{deterministic} suspensions in general domains by constructing complex particle shapes out of agglomerates of minimally resolved spheres/blobs, termed \emph{rigid multiblobs}. Both \cite{BrownianBlobs} and \cite{RigidMultiblobs} employ the Immersed Boundary (IB) method to handle the fluid-particle coupling. IB methods provide a low-accuracy but inexpensive and flexible alternative to body-fitted grid-based methods since no remeshing is required as particles move around in the domain.
In \cite{BrownianMultiblobSuspensions}, some of us presented an efficient temporal integration scheme to simulate the dynamics of Brownian suspensions with many rigid particles of arbitrary shape confined above a single no-slip wall. This method relies on the simple geometry of the physical domain for which an explicit form for the hydrodynamic mobility operator is available. Here we present an amalgamation of these past approaches: we develop a generalization of the temporal integration of \cite{BrownianMultiblobSuspensions} that fits the IB framework used in \cite{RigidMultiblobs} to handle the hydrodynamic interactions including with boundaries, and use the fluctuating hydrodynamics approach proposed in \cite{BrownianBlobs} to account for Brownian motion.
We develop a novel Split Euler--Maruyama (SEM) temporal integration scheme to capture the stochastic drift which strongly affects even a single particle. The SEM scheme modifies the preconditioned Krylov method of \cite{RigidMultiblobs} to maintain its linear scaling but includes the necessary stochastic contributions to the rigid-body dynamics.


In \cite{Bohlein2011} the authors experimentally observed soliton wave patterns in a driven colloidal monolayer moving above a bottom substrate. The bottom wall is patterned with a periodic potential meant to mimic surface roughness. The monolayer is forced into quasi-2D confinement by laser-induced forces, and driven by the flow generated by moving the bottom wall. Brownian motion is crucial in activating the transitions of the colloids between the minima of the patterned potential, and must be captured accurately to resolve the dynamics of the monolayer. In section \ref{lattice}, we use the RB-FIB method to numerically investigate a modified version of the experiment performed in \cite{Bohlein2011} where we confine the monolayer in a thin microchannel. We observe novel wave patters in the colloidal monolayer which emerge due to the physical confinement. While simulations of this type have been performed \cite{ColloidSheetFriction,LatticeNoHydro}, our work is, to our knowledge, the first which includes an accurate treatment of the hydrodynamics.

The rest of this work is organized as follows. In section \ref{background} we give the continuous formulation of the problem and introduce some relevant notation. In section \ref{DF} we formulate the spatial discretization of the continuum equations. In section \ref{tint} we introduce the SEM scheme as an efficient temporal discretization that maintains discrete fluctuation dissipation balance. To numerically validate our scheme, we consider several test cases. Appendix \ref{sec:boom} considers a boomerang shaped particle in a slit channel, and confirms that the RB-FIB method is first order weakly accurate for expectations with respect to the Gibbs-Boltzmann equilibrium distribution. Section \ref{twoSphere} considers two spherical particles trapped in a tight cuboidal box to examine the effect of spatial resolution on dynamic statistics. In section \ref{lattice} we study the transition from static to dynamic friction for a suspension of many spherical colloids  confined to a quasi--two dimensional slit channel, and hydrodynamically driven across a periodic substrate potential by translating the microchannel with constant velocity. We conclude with a summary and discussion of future directions in section \ref{sec:conc}

\section{Continuum Formulation} \label{background}

The fluctuating Stokes equations in a physical domain $\Omega$ can be written as \cite{FluctHydroNonEq_Book}
\begin{align} \label{FHD}
\rho \partial_{t} \V{v} &= \grad \cdot \V{\Sigma} + \V{g} = \grad \cdot \left( \V{\sigma} + \sqrt{2 k_{B} T \eta} \sM{Z} \right) + \V{g}, \\
\grad \cdot \V{v} &= 0, \nonumber
\end{align}
where the fluid has density $\rho$, shear viscosity $\eta$, and temperature $T$, all of which we take to be constant in this work. Denoting Cartesian coordinates with $\V{x} \in \Omega$ and time with $t$,  $\V{g}\left(\V{x},t \right)$ represents a fluid body force and $\sigma = -\pi \M{I} + \eta \left( \grad \V{v} + \grad \V{v}^{T}\right)$ is the dissipative component of the fluid stress tensor, where $\V{v}\left(\V{x},t \right) $ is the fluid velocity, and $\pi\left(\V{x},t \right)$ is the pressure. The \emph{stochastic stress tensor} $\sqrt{2 k_{B} T \eta} \sM{Z}$ accounts for the fluctuating contribution to the fluid stress and ensures fluctuation dissipation balance,  where $\sM{Z}\left(\V{x},t \right)$ is a symmetric random Gaussian tensor whose components are delta correlated in space and time,
\[
 \av{\mathcal{Z}_{ij}(\V{x},t)\mathcal{Z}_{kl}(\V{x}',t')} = \left( \delta_{ik} \delta_{jl} + \delta_{il} \delta_{jk} \right) \delta(t-t')\delta(\V{x} - \V{x}').
\]

We consider $N_b$ arbitrarily shaped rigid particles (bodies) $\mathcal{B}_p$, $1 \leq p \leq N_b$, suspended in the fluctuating Stokesian fluid. Because the particles are treated as rigid, their position in space can be completely described by the Cartesian location $\V{q}_p(t)$ of a representative tracking point, and an orientation $\V{\theta}_p(t)$ relative to a reference configuration. The bulk of our discussion will be independent of how one wishes to describe the orientation of the particles, however, in practice, we  use unit quaternions, as described in \cite{BrownianMultiBlobs}. The composite configuration of particle $p$ will be denoted by $\V{Q}_p = \left[\V{q}_p, \V{\theta}_p \right]$.

Rigid body $p$ moves with translational velocity $\V{u}_p = d\V{q}_p/dt$ and rotates with angular velocity $\V{\omega}_p$ around the tracking point. The composite velocity $\V{U}_p = \left[\V{u}_p, \V{\omega}_p \right]$ can be used to express the velocity of an arbitrary point $\V{r} \in \partial \mathcal{B}_p$ on the surface of the particle to give the \emph{no-slip condition}
\begin{equation}\label{rigidBC}
\V{v}\left( \V{r} \right) = \V{u}_{p} + \left( \V{r} - \V{q}_{p} \right) \times \V{\omega}_{p} + \breve{\V{u}}_p.
\end{equation}
Here $\breve{\V{u}}_p$ is an apparent slip between the fluid and the particle surface which may be freely prescribed up to the condition that the integral of the slip velocity vanish over the surface of the particle \cite{FBIM}. The free slip velocity can be used to account for active layers formed by, say, electrokinetic flows or beating flagella. 

Newton's second law gives the acceleration of particle $p$ in terms of the applied force $\V{f}_{p}$ and the applied torque $\V{\tau}_{p}$, 
\begin{align} 
m_{p} \frac{\partial \V{u}_p}{\partial t} &= \V{f}_{p} - \displaystyle \int_{\partial \mathcal{B}_{p}} \left(\V{\Sigma} \cdot \V{n} \right) d A( \V{r}), \label{CFTB1} \\
\M{I}_{p} \frac{\partial \V{\omega}_p}{\partial t} &= \V{\tau}_{p} -  \displaystyle \int_{\partial \mathcal{B}_p}  \left( \V{r} - \V{q}_{p} \right) \times \left(\V{\Sigma} \cdot \V{n} \right)  d A( \V{r}), \label{CFTB2}
\end{align}
where $\V{n}\left( \V{r}\right) $ is the outward pointing surface normal vector, and $m_{p}$ and $\M{I}_{p}$ are the particles mass and angular inertia tensor respectively.

In this work we are interested in the \emph{overdamped} or steady Stokes limit of the dynamics, which is the one relevant for colloidal suspensions due to the very large Schmidt numbers and very small Reynolds numbers. In the \emph{absence} of thermal fluctuations, the steady Stokes or inertia-less limit of equations \eqref{CFTB1}-\eqref{CFTB2}, \eqref{rigidBC}, and \eqref{FHD} is taken by simply deleting the inertial terms to obtain
\begin{align}
- \eta \grad^2 \V{v} + \grad \pi = \V{g}, \ &\grad \cdot \V{v} = 0 \label{DetC1}  \text{ in } \Omega \setminus \left\lbrace \cup_p \mathcal{B}_p \right\rbrace,  \\
\V{v}(\V{r}) &= \V{u}_{p} + \left( \V{r} - \V{q}_{p} \right) \times \V{\omega}_{p} + \breve{\V{u}}: \hspace{0.5cm}  \forall p, \forall \V{r} \in \partial \mathcal{B}_{p}, \label{DetC2} \\
\V{f}_{p} = \displaystyle \int_{\partial \mathcal{B}_{p}} \V{\sigma} \cdot \V{n} \left(\V{r}\right) d A( \V{r}), \ \ \V{\tau}_{p} &=  \displaystyle \int_{\partial \mathcal{B}_p}  \left( \V{r} - \V{q}_{p} \right) \times \left( \V{\sigma} \cdot \V{n} \right) \left(\V{r}\right)  d A( \V{r}):\hspace{0.5cm} \forall p. \label{DetC3}
\end{align}
The solution to the linear system of equations \eqref{DetC1}--\eqref{DetC3} can be expressed using a symmetric, positive semi-definite \emph{body mobility matrix} $\N(\V{Q})$ which is a function of the composite configuration $\V{Q} = \left[\V{q}_{1}, \V{\theta}_{1}, \ldots ,\V{q}_{N_b}, \V{\theta}_{N_b} \right]$. The body mobility matrix acts on the external forces and torques applied to each particle $\V{F} = \left[ \V{f}_{1}, \V{\tau}_{1}, \ldots ,\V{f}_{N_b}, \V{\tau}_{N_b} \right]$ to produce a composite vector $\V{U} = \left[ \V{u}_{1}, \V{\omega}_{1}, \ldots, \V{u}_{N_b}, \V{\omega}_{N_b} \right]$ of the rigid body velocities of the particles, $\V{U}=\N \V{F}$. 

If we \emph{include} the thermal fluctuations, the overdamped limit of equations \eqref{CFTB1}-\eqref{CFTB2}, \eqref{rigidBC}, and \eqref{FHD} is the overdamped Langevin equation \cite{LangevinDynamics_Theory,FBIM}
\begin{align} 
\V{U} &= \N \V{F} + \sqrt{2 k_{B} T} \ \N^{1/2} \diamond \sM{W} \label{LangevinNd} \\
 &= \N \V{F} + k_{B} T \left(\partial_{\V{Q}} \cdot \N \right) + \sqrt{2 k_{B} T} \ \N^{1/2} \sM{W} \label{LangevinN}
\end{align}
where $\diamond$ is the kinetic stochastic product \cite{KineticStochasticIntegral_Ottinger}, and $\sM{W}(t)$ is a collection of independent Wiener processes. Equation \eqref{LangevinN} is the conversion of \eqref{LangevinNd} to Ito form, by introducing the term $k_{B} T \left(\partial_{\V{Q}} \cdot \N \right)$ which we call the `thermal' or `stochastic' drift \footnote{The precise mathematical interpretation of this notation when $\V{Q}$ includes particle orientations expressed in terms of quaternions is explained in \cite{BrownianMultiBlobs}.}. Fluctuation dissipation balance is maintained in \eqref{LangevinN} through the random vector $\N^{1/2} \sM{W}$ such that $ \N^{1/2} \left(\N^{1/2}\right)^{\star} = \N$, where star denotes an $L_2$ adjoint (i.e., a conjugate transpose for matrices). Note that the dynamics of the particle configurations $d\V{Q}/dt$ can be directly expressed in terms of $\V{U}$ using the quaternion representation of the particle orientations, as discussed in \cite{BrownianMultiBlobs}.

Numerical methods for temporal integration of \eqref{LangevinN} in time are discussed in \cite{BrownianMultiblobSuspensions}. These methods require an efficient method to generate both the deterministic and Brownian velocities of the particles. Specifically, efficient Brownian dynamics for rigid bodies requires efficiently computing an Euler-Maruyama approximation of the apparent linear and angular velocities over a time step of duration $\D{t}$,
\begin{equation} \label{UEMdef}
\V{U}_{\text{EM}} = \N \V{F} + \sqrt{\frac{2 k_{B} T}{\D{t}}} \ \N^{1/2} \V{W},
\end{equation}
where $\V{W}$ is a collection of independent standard Gaussian random variables. Much of the strength and flexibility of the RB-FIB method we introduce in this work comes from the fact that we compute $\V{U}_{\text{EM}}$ by discretizing a semi--continuum formulation of the overdamped particle dynamics (temporarily neglecting the thermal drift) instead of relying on explicit representations of Greens functions as in \cite{BrownianMultiblobSuspensions,FBIM}. The missing stochastic drift term in \eqref{LangevinN} involving $\partial_{\V{Q}} \cdot \N$ can be obtained in expectation by adding a correction to the Euler-Maruyuama method, as we explain in detail in Section \ref{tint}. 

Temporarily neglecting the terms that contribute to the stochastic drift, the overdamped limit of equations \eqref{CFTB1}-\eqref{CFTB2}, \eqref{rigidBC}, and \eqref{FHD} can be obtained quite simply by deleting all of the inertial terms and replacing the space-time white noise field $\sM{Z}(\V{x},t)$ with a white-in-space random Gaussian field $\V{Z}(\V{x})/\sqrt{\D{t}}$. In the immersed boundary approach we use here, we extend the fluid equation over the whole domain, including inside the bodies, since \eqref{rigidBC} is satisfied for all $\V{r}\in \mathcal{B}_p$ including the rigidly moving interior \cite{RigidMultiblobs}. This gives a system of semi--continuum linear equations for $\V{U}_{\text{EM}} = \left[ \V{u}^{\text{EM}}_{1}, \V{\omega}^{\text{EM}}_{1}, \ldots  \V{u}^{\text{EM}}_{N_b}, \V{\omega}^{\text{EM}}_{N_b} \right]$, 
\begin{align} 
-\eta \grad^2 \V{v} + \grad \pi =  \sqrt{\frac{2 k_{B} T \eta}{\D{t}}}& \grad \cdot \V{Z} + \displaystyle \sum_p \displaystyle \int_{\partial \mathcal{B}_{p}} \delta \left(\V{x} -\V{r} \right) \V{\lambda}\left(\V{r}\right) d A( \V{r})  + \V{g}: \hspace{0.5cm} \forall \V{x} \in \Omega \label{LangevinC1} \\
\grad \cdot \V{v} &= 0 : \hspace{0.5cm} \forall \V{x} \in \Omega  \label{LangevinC2} \\
\displaystyle \int_{\Omega} \delta \left(\V{x} -\V{r} \right) \V{v}\left( \V{x} \right) d V( \V{x}) &= \V{u}^{\text{EM}}_{p} + \left( \V{r} - \V{q}_{p} \right) \times \V{\omega}^{\text{EM}}_{p} + \breve{\V{u}}: \hspace{0.5cm}  \forall p, \forall \V{r} \in \partial \mathcal{B}_{p}, \label{LangevinC3} \\
\V{f}_{p} = \displaystyle \int_{\partial \mathcal{B}_{p}} \V{\lambda}\left(\V{r}\right) d A( \V{r}), \ \ \V{\tau}_{p} &=  \displaystyle \int_{\partial \mathcal{B}_p}  \left( \V{r} - \V{q}_{p} \right) \times \V{\lambda}\left(\V{r}\right)  d A( \V{r}):\hspace{0.5cm} \forall p. \label{LangevinC4}
\end{align}
We do not add a superscript `EM' to the velocity and pressure here with the understanding that in the overdamped limit they are just auxiliary variables used to obtain the motion of the particles. Here $\V{\lambda}$ is the jump in the fluid stress across the boundary of the particles, $\V{\lambda} \equiv \llbracket \V{\Sigma} \cdot \V{n} \rrbracket$, which can be simply identified as the traction force when $\breve{\V{u}}=\V{0}$; see Appendix A in \cite{RigidMultiblobs} for an explanation why the same formulation works even when the apparent slip $\breve{\V{u}}$ is nonzero. The use of a Dirac--delta distribution to restrict quantities to their values on the surface of the bodies $\V{r} \in \partial \mathcal{B}_{p}$ is the basis for the immersed boundary spatial discretization described next.

\section{Discrete Formulation} \label{DF}

In this section we describe an efficient and robust means of discretizing the problem formulated in section \ref{background}. We will use standard finite difference and immersed boundary methods to construct matrix discretizations of the differential and integral operators appearing in \eqref{LangevinC1}--\eqref{LangevinC4}. We will briefly review efficient methods to solve the large-scale linear system which arises in the discretized equations; details can be found in \cite{RigidIBM,RigidMultiblobs}.

\subsection{Discrete Fluctuating Stokes Equations}

To begin discretizing equations \eqref{LangevinC1}--\eqref{LangevinC4} we consider the discretization of the fluctuating Stokes equations without any suspended particles. Specifically, we will temporarily ignore the effect of the rigid bodies on the fluid in equations \eqref{LangevinC1}-\eqref{LangevinC2} by setting $\V{\lambda} = 0$. Importantly, all of the discussion in this section will remain agnostic to the choice of physical boundary conditions on $\Omega$.

We discretize the Stokes equations on a regular Cartesian `Eulerian' grid of cells with volume $\D{V}=h^d$, where $h$  is the grid spacing and $d$ is the space dimension. Velocity variables are staggered on the faces of the grid relative to the cell-centered pressure variables. The infinite dimensional white noise field $\V{Z}$ is spatially discretized as $\V{W}$, a collection of random Gaussian variables generated on the faces and nodes of the fluid grid, as described in \cite{LLNS_Staggered}. We use staggered discretizations of the vector divergence $\Dm$ and scalar gradient $\Gm$ operators that obey the adjoint relation $\Gm = -{\Dm}^{\star}$. We also define discrete tensor divergence $\DTm$ and vector gradient $\GTm=-\DTm^{\star}$ operators, which account for the imposed boundary conditions on $\partial\Omega$. The discrete scalar Laplacian is $\M{L} = \Dm \Gm$ and the discrete vector Laplacian is $\Lm = \DTm \GTm$ in order to satisfy a discrete fluctuation dissipation balance principle \cite{LLNS_Staggered}.

With $\V{\lambda} = 0$, we discretize equations \eqref{LangevinC1}-\eqref{LangevinC2} as
\begin{align} \label{DFHD}
  - \eta \Lm \V{v} + \Gm \pi &= \V{g} + \left(\frac{2 k_{B} T \eta}{\D{V} \D{t}}\right)^{1/2} \DTm \M{W} = \V{f}, \\
  \Dm \V{v} &= 0. \nonumber
\end{align}
This maintains discrete fluctuation dissipation balance even in the presence of physical boundaries. It is convenient at this point to introduce the symmetric positive-semidefinite discrete Stokes solution operator 
$$-\sM{L}^{-1} = \frac{1}{\eta}\left(\Lm^{-1}-\Lm^{-1}\M G\left(\M D\M \Lm^{-1}\M G\right)^{-1}\M D\M \Lm^{-1}\right),$$
such that the solution to \eqref{DFHD} can be written as
$\V{v} = \sM{L}^{-1} \V{f}.$

\subsection{Immersed Boundary Method for Rigid Bodies}

To mediate the fluid-structure interaction we use a discrete approximation $\delta_h$ to the Dirac delta distribution appearing in \eqref{LangevinC1}--\eqref{LangevinC4}. In the results presented here $\delta_h$ is the 6-point kernel developed in \cite{New6ptKernel}. While the discretization on the grid introduces numerical artifacts in general \cite{IBM_PeskinReview}, the 6-point kernel we use here was specifically designed with grid invariance and smoothness in mind \cite{New6ptKernel}. \deleted{There are a number of other common choices for the kernel, $\delta_h$, used in $\J$ and $\S$. For instance, the Gaussian kernel of the force coupling method \cite{FluctuatingFCM_DC} which is isotropic and generalizes to non-spherical blob shapes, or the classical and well documented 4 point Immersed boundary kernel described by Peskin \cite{IBM_PeskinReview}.}

Throughout the rest of this work, we will represent a body $\mathcal{B}_p$ as a rigid agglomerate of markers or \emph{blobs} with positions $\V{r}^{p}_{i} \in \partial\mathcal{B}_p$ and we refer to the collection of these points as the `Lagrangian' grid. The ideal spacing between the blobs $s \sim h$ is related to the meshwidth $h$ used in discretizing the fluid equations \eqref{DFHD}, as discussed in detail in section IV of \cite{RigidMultiblobs}. We discretize $\V{\lambda}$ on the Lagrangian grid as a collection of force vectors $\V{\lambda}_{i}^{p} \approx \V{\lambda}\left( \V{r}^{p}_{i} \right) \D{A}\left( \V{r}^{p}_{i} \right)$. It is important to note that the discrete $\V{\lambda}_{i}^{p}$ has units of force rather than force density as the continuum $ \V{\lambda}\left( \V{r} \right)$. Recall that the fluid velocity $\V{v}$ is defined on the centers of the faces $\V{x}_\alpha$ of the Eulerian grid.

To discretize equations \eqref{LangevinC1}--\eqref{LangevinC4} we use simple trapezoidal rule quadratures to approximate the integrals as appropriate sums over Eulerian or Lagrangian grid points. This leads us to define the \emph{spreading operator} $\S$ and the \emph{interpolation operator} $\J$ as
\begin{align}
\left( \J \V{v} \right)_i^p &= \displaystyle \sum_{\V{x}_\alpha \in \Omega } \delta_{h}\left(\V{x}_{\alpha} - \V{r}^p_{i} \right) \V{v}\left( \V{x}_{\alpha} \right) \approx \displaystyle \int_{\Omega} \delta \left( \V{x} - \V{r}_i^p \right) \V{v}\left( \V{x} \right) d V(\V{x}), \\
\left( \S \V{\lambda} \right)_\alpha &= \frac{1}{\D{V}} \displaystyle \sum_p \sum_{\V{r}^{p}_{i}  } \delta_{h}\left( \V{x}_\alpha - \V{r}^p_{i} \right) \V{\lambda}_{i}^p \approx  \displaystyle \int_{\partial \mathcal{B}^{p}} \delta \left( \V{x}_{\alpha} - \V{r} \right) \V{\lambda}\left( \V{r} \right) d A(\V{r}).
\end{align}

It is important to note that $\J$ and $\S$ satisfy the adjoint relation $\J = \D{V} \S^{\star}$. This property ensures discrete conservation of energy \cite{IBM_PeskinReview} as well as discrete fluctuation dissipation balance. The definitions of $\S$ and $\J$ are modified when the support of the kernel overlaps with a physical boundary of $\Omega$. Appropriate ghost points across the physical boundary are used which ensures the effects of the boundary are incorporated into the spreading and interpolation operations while preserving the adjoint property, as described in Appendix D of \cite{RigidIBM}.

We discretize the integrals giving the total force and torque in \eqref{LangevinC4} using simple trapezoidal quadrature to define the geometric matrix $\K(\V{Q})$ \cite{RigidMultiblobs_Swan},
\begin{align}
\left(\K  \V{U} \right)_i^p &= \V{u}_{p} + \left( \V{r}^{p}_i - \V{q}_{p} \right) \times \V{\omega}_{p},\\
\left(\K^T  \V{\lambda} \right)_p &= \begin{bmatrix}
\displaystyle \sum_{\V{r}^{p}_{i} } \V{\lambda}^p_i \\
\displaystyle \sum_{\V{r}^{p}_{i} }  \left( \V{r}^{p}_i - \V{q}_{p} \right) \times \V{\lambda}^p_i
\end{bmatrix}
\approx \begin{bmatrix}
\displaystyle \int_{\partial \mathcal{B}_{p}} \V{\lambda}\left(\V{r}\right) d A(\V{r}) \\
\displaystyle \int_{\partial \mathcal{B}_p}  \left( \V{r} - \V{q}_{p} \right) \times \V{\lambda}\left(\V{r}\right)  d A(\V{r})
\end{bmatrix}.
\end{align}

\subsection{The Discrete System}

We can now compactly state the spatially discretized system \eqref{LangevinC1}--\eqref{LangevinC4} as
\begin{align} 
- \eta \Lm \V{v} + \Gm \pi &= \V{g} + \S \V{\lambda} + \left(\frac{2 k_{B} T \eta}{\D{V} \D{t}}\right)^{1/2} \DTm \M{W}, \label{LangevinD1} \\
  \Dm \V{v} &= 0, \label{LangevinD2} \\
\J \V{v} &= \K \V{U}_{\text{EM}} + \slip, \label{LangevinD3} \\
\K^{T} \V{\lambda} &= \V{F}. \label{LangevinD4}
\end{align}
For simplicity in the following discussion, we will take the fluid body force $\V{g}=\V{0}$. 

Using $\S$ and $\J$ we may define a regularized, symmetric, positive semi-definite blob-blob mobility matrix $\Mob = \J  \L^{-1} \S$, where we recall that $\L^{-1}$ denotes the discrete Stokes solution operator. The block $\Mob_{ij}$ gives the pairwise mobility matrix between two blobs $\V{r}_{i}$ and $\V{r}_{j}$ \cite{BrownianBlobs,RigidIBM,RigidMultiblobs},
\begin{equation} \label{MobDefn}
\Mob_{ij} \approx \displaystyle \iint_{\Omega \times \Omega} \delta_{h}\left(\V{x} - \V{r}_{i} \right) \Set{G} \left( \V{x},\V{y} \right) \delta_{h}\left( \V{y} - \V{r}_{j} \right) d V(\V{y}) d V(\V{x})
\end{equation}
where $\Set{G}$ is the Green's function for the Stokes equations in $\Omega$ with the specified boundary conditions. For two markers/blobs that are sufficiently far apart in a sufficiently large domain, $\Mob_{ij}$ approximates the mobility for a pair of spheres of radius \footnote{Specifically, $a=1.47 h$ for the 6-point kernel used in this work \cite{RigidIBM,RigidMultiblobs}.} $a \sim h$. When the Green's function is available analytically, $\Mob$ can be computed explicitly; here we handle more general boundary conditions by solving the discretized steady Stokes equations numerically to compute the action of $\Set{G}$.

We may use the definition of $\Mob$ to eliminate the velocity and pressure from \eqref{LangevinD1}--\eqref{LangevinD4} to obtain the reduced linear system
\begin{align} 
\Mob \V{\lambda} &= \K \V{U}_{\text{EM}} + \slip -\sqrt{\frac{2 k_{B} T}{\D{t}}} \Mob^{1/2} \V{W}, \label{LangevinM1} \\
\K^{T} \V{\lambda} &= \V{F}. \label{LangevinM2}
\end{align}
Here we can identify the matrix $\Mob^{1/2} = \sqrt{\eta/\D{V}} \J \L^{-1} \DTm$ as was done in \cite{BrownianBlobs}, such that $\Mob^{1/2} \left(\Mob^{1/2}\right)^{\star} = \Mob$. Equation \eqref{LangevinM1}--\eqref{LangevinM2} is identical to equation (9) in \cite{BrownianMultiblobSuspensions} -- the only difference here is that we do not explicitly have access to $\Mob$ because we do not necessarily know the Green's function $\Set{G}$. Following \cite{BrownianMultiblobSuspensions}, the solution of \eqref{LangevinM1}--\eqref{LangevinM2}  can be written in terms of the body mobility matrix $\N = \left( \K^{T} \Mob^{-1} \K \right)^{-1}$,
\begin{equation} \label{incompleteN}
\V{U}_{\text{EM}} = \N \V{F} - \N \K^{T} \Mob^{-1} \slip + \sqrt{\frac{2 k_{B} T}{\D{t}}} \N^{1/2} \M{W},
\end{equation}
where we have identified $\N^{1/2} \equiv \N \K^T \Mob^{-1} \Mob^{1/2}$, such that $\N^{1/2} \left(\N^{1/2}\right)^{\star}=\N$. 

We have now shown that we can compute $\V{U}_{\text{EM}}$ given by \eqref{incompleteN} efficiently by solving
the linear system \eqref{LangevinC1}--\eqref{LangevinC4}, which can be done with a complexity linear in the number of particles thanks to the preconditioned GMRES solver developed in \cite{RigidMultiblobs}.
All the remains to perform Brownian Dynamics is an efficient means of computing the stochastic drift term $k_{B} T \left(\partial_{\V{Q}} \cdot \N \right)$ in \eqref{LangevinN}, as we discuss next.

\section{Temporal Integration} \label{tint}

In this section, we develop a temporal integration scheme for equation \eqref{LangevinN} which efficiently captures the contribution from the stochastic drift term $k_{B} T \left(\partial_{\V{Q}} \cdot \N \right)$. Specifically, we use \emph{random finite differences} (RFDs) \cite{BrownianBlobs,BrownianMultiBlobs,MagneticRollers} to compute terms that will be included on the right hand side of equations \eqref{LangevinD1}--\eqref{LangevinD4} to account for the stochastic drift in expectation.
The algorithm we develop here requires the solution of an additional linear system similar to \eqref{LangevinD1}--\eqref{LangevinD4} each time step in order to capture the stochastic drift. This is still more efficient than the classical Fixman scheme which requires solving a resistance problem, which is a lot more expensive than solving a mobility problem when iterative methods are used \cite{RigidMultiblobs,libStokes}.

\subsection{Random Finite Differences}

Solving equations \eqref{LangevinD1}--\eqref{LangevinD4} without any modifications yields an efficient means of computing $\V{U}_{\text{EM}}$ and all that remains to simulate equation \eqref{LangevinN} is a means of computing $k_{B} T \left(\partial_{\V{Q}} \cdot \N \right)$. In past work \cite{BrownianBlobs,BrownianMultiBlobs,MagneticRollers}, some of us have proposed to use random finite differences (RFD) to generate this term as follows.
Consider two Gaussian random vectors  $\Delta \V{P}$ and $\Delta \V{Q}$, such that $\av{\Delta \V{P}} = \av{\Delta \V{Q}} = 0$ and $\av{\Delta \V{P} \Delta \V{Q}^{T}} = \M{I}$. For an arbitrary matrix $\sM{R}\left(\V{Q}\right)$, it holds that
\begin{align} \label{RFD}
\partial_{\V{Q}} \cdot \V{R} \left( \V{Q} \right) &=  \lim_{\delta \rightarrow 0} \frac{1}{\delta}\av{\V{R} \left( \V{Q} + \delta \Delta \V{Q} \right) - \V{R} \left( \V{Q} \right)} \Delta \V{P}\\
&= \lim_{\delta \rightarrow 0} \frac{1}{\delta}\av{\V{R} \left( \V{Q} + \frac{\delta}{2} \Delta \V{Q} \right) - \V{R} \left( \V{Q} - \frac{\delta}{2} \Delta \V{Q} \right)} \Delta \V{P},
\end{align}
where $\av{\cdot}$ denotes an average over realizations of the random vectors.

The limits in equations \eqref{RFD} may be discretely approximated by simply choosing a small value for $\delta$ at the cost of introducing a truncation error of $\mathcal{O}\left( \delta \right)$ if the one--sided difference is used (first line) or $\mathcal{O}\left( \delta^2 \right)$ if the centered difference is used (second line). The value of $\delta$ should be chosen to balance the magnitude of the truncation errors in the RFD with any numerical error associated with the application of (multiplication with) $\V{R}$.  Specifically, we take $\delta=\epsilon^{1/2}$ for the one-sided difference and $\delta=\epsilon^{1/3}$ for the centered difference, where $\epsilon$ is the relative error with which matrix-vector products with $\V{R}$ are computed.

A simple modification of the Euler--Maruyama scheme is to use a RFD with $\sM{R} \equiv \N$ to account for the stochastic drift \cite{BrownianMultiBlobs}. Since application of $\N$ requires solving \eqref{LangevinD1}--\eqref{LangevinD4} iteratively with some loose relative tolerance $\epsilon$, the required value of $\delta$ would be relatively large especially for the one-sided difference. Using the two-sided difference requires solving two additional mobility problems per time step, which is quite expensive. We now propose an alternative approach.

\subsection{The Split--Euler--Maruyama (SEM) Scheme}

Our goal to design a means of computing the stochastic drift with as few linear solves as possible. In our prior work \cite{BrownianMultiblobSuspensions}, we accomplished this by expanding $\partial_{\V{Q}} \cdot \N$ using the chain rule,
\begin{align} \label{RFDsplit}
&\partial_{\V{Q}} \cdot \N = \partial_{\V{Q}} \cdot \left(\K^{T} \Mob^{-1} \K \right)^{-1} = \nonumber \\
   &-\N  \left( \partial_{\V{Q}} \K^{T} \right) \colon \Mob^{-1} \K \N - \N \K^{T} \Mob^{-1} \left( \partial_{\V{Q}} \Mob \right) \colon \Mob^{-1} \K \N + \N \K^{T} \Mob^{-1} \left( \partial_{\V{Q}} \K \right) \colon \N.
\end{align}
In this work we do not have explicit access to $\Mob$ so we carry this expansion one step further as done in \cite{BrownianBlobs},
\begin{align} \label{divM}
\partial_{\V{Q}} \cdot \Mob = \partial_{\V{Q}} \cdot \left(\J \L^{-1}\S\right) =  \left( \partial_{\V{Q}} \J \right) \colon \sM{L}^{-1} \S  + \J \sM{L}^{-1} \left( \partial_{\V{Q}} \cdot \S \right).
\end{align}

Our aim is to generate each term in \eqref{RFDsplit} and \eqref{divM} through a separate RFD on $\J$, $\S$, $\K$, or $\K^{T}$. This is particularly advantageous as these operators can all be applied efficiently to within roundoff tolerance in linear time, without requiring linear solvers.

The Euler--Maruyama--Traction (EMT) scheme proposed in \cite{BrownianMultiblobSuspensions} can be adapted to the present context to give the \emph{Split--Euler--Maruyama} (SEM) scheme outlined in Algorithm \ref{alg:sem}. In the first step of this algorithm we generate random forces and torques on each body 
\begin{equation}
\V{W}^{\text{FT}} = k_{B} T \begin{bmatrix}
          \frac{1}{L_p}\V{W}_{p}^{\text{f}} \\
          \V{W}_{p}^{\tau}
\end{bmatrix},
\end{equation}
where $\V{W}_{p}^{\text{f}/\tau}$ are standard Gaussian random variables generated independently for each body $p$. Here  $L_p$ is a length scale for body $p$, which we take as the maximum pairwise distance between blobs on a body. In step \ref{step:solve1} of Algorithm \ref{alg:sem} we solve a mobility problem with $\V{W}^{\text{FT}}$ as the applied force/torque on each body, to obtain the random variables
\begin{align} \label{ulvRFD}
\V{U}^{RFD} &= \N \V{W}^{\text{FT}}, \\
{\V{\lambda}}^{RFD} &= \Mob^{-1}  \K \N \V{W}^{\text{FT}}, \\
\V{v}^{RFD} &= - \L^{-1} \S \Mob^{-1}  \K \N \V{W}^{\text{FT}}.
\end{align}

Defining a random translational and rotational displacement for body $p$
\begin{equation}
\Delta \V{Q}_p = \begin{bmatrix}
          L_p \V{W}_{p}^{\text{f}} \\
          \V{W}_{p}^{\tau}
\end{bmatrix},
\end{equation}
gives two randomly diplaced positions \footnote{Here for simplicity of notation we use addition to denote a random rotation of the body by an oriented angle $(\delta/2) \V{W}_{p}^{\tau}$ even though in practice this is realized as a quaternion multiplication in three dimensions \cite{BrownianMultiBlobs}.} of body $p$,
\begin{equation}
\V{Q}_p^{\pm} = \V{Q}_p \pm \frac{\delta}{2} \Delta \V{Q}_p.
\end{equation}
We can produce the desired drift term through the following random finite differences on the matrices $\K^{T}$ and $\K$:
\begin{align} 
 \V{D}^{\K^{T}} &= \frac{1}{\delta} \left[ \K^{T} \left(\V{Q}^{+}\right)  - \K^{T} \left(\V{Q}^{-}\right) \right] {\V{\lambda}}^{RFD} \approx \left( \partial_{\V{Q}} \K^{T} \right) \colon \Mob^{-1} \K \N \left(\V{W}^{\text{FT}} \Delta \V{Q}^{T} \right), \label{allDemBoysa} \\
 \V{D}^{\K} &= \frac{1}{\delta} \left[ \K \left(\V{Q}^{+}\right)  - \K \left(\V{Q}^{-}\right) \right] \V{U}^{RFD} \approx \left( \partial_{\V{Q}} \K \right) \colon \N \left(\V{W}^{\text{FT}} \Delta \V{Q}^{T} \right), \label{allDemBoysb}
\end{align}
which are analogous to the quantities $\V{D}^{F}$ and $\V{D}^{S}$ computed in Algorithm 1 of \cite{BrownianMultiblobSuspensions}.

However, the RFD performed directly on $\Mob$ to compute $\V{D}^{S}$ in the EMT scheme (see Algorithm 1 of \cite{BrownianMultiblobSuspensions}) is computed in the SEM scheme using \eqref{divM} as a sum of RFDs on the interpolation and spreading operators,
\begin{align} 
 \V{D}^{\J} &= \frac{1}{\delta} \left[ \J \left(\V{Q}^{+}\right)  - \J \left(\V{Q}^{-}\right) \right] \V{v}^{RFD} \approx \left( \partial_{\V{Q}} \J \right) \colon \sM{L}^{-1} \S \Mob^{-1}  \K \N \left(\V{W}^{\text{FT}} \Delta \V{Q}^{T} \right). \label{allDemBoys2a}\\
 \V{D}^{\S} &= \frac{1}{\delta} \J \L^{-1} \left[\S \left(\V{Q}^{+}\right) - \S \left(\V{Q}^{-}\right) \right] {\V{\lambda}}^{RFD} \approx \J \sM{L}^{-1} \left( \partial_{\V{Q}} \S \right) \colon \Mob^{-1} \K \N \left(\V{W}^{\text{FT}} \Delta \V{Q}^{T} \right). \label{allDemBoys2b}
\end{align}
Note that the computation of $\V{D}^{\S}$ in step \ref{alg:uncon} of Algorithm \ref{alg:sem} requires an additional application of $\L^{-1}$ and therefore an additional unconstrained Stokes solve. This does not add much additional complexity to the computation because unconstrained (fluid only) Stokes systems of the form \eqref{DFHD} have fewer degrees of freedom and are far better conditioned than constrained (fluid + rigid bodies) systems of the form \eqref{LangevinD1}--\eqref{LangevinD4} \cite{RigidMultiblobs}.

To produce the correct drift term, we add $\V{D}^{\K} - \V{D}^{\J} - \V{D}^{\S}$ as a random slip and add $\V{D}^{\K^{T}}$ as a random force on the right hand side of the linear system in step \ref{step:solve2} of Algorithm \ref{alg:sem}. This generates an additional contribution to the velocity of the rigid particles, $\V{U}^{n} = \V{U}_{\text{EM}} + \V{U}_{\text{Drift}}$, where
\begin{align} \label{DriDef}
 \V{U}_{\text{Drift}} &= \N \V{D}^{\K^{T}} + \N \K^{T} \Mob^{-1} \left( -\V{D}^{\K} + \V{D}^{\J} + \V{D}^{\S} \right) \nonumber \\
 &=\partial_{\V{Q}} \N  \colon \left(\V{W}^{\text{FT}} \Delta \V{Q}^{T} \right).
\end{align}
We used equations \eqref{allDemBoysa}--\eqref{allDemBoys2b} as well as equation \eqref{RFDsplit} to simplify from the first to the second line in \eqref{DriDef}. On average (i.e. in expectation) this will produce the desired drift term,
\begin{equation}
\av{\V{U}_{\text{Drift}}} = (k_B T) \partial_{\V{Q}} \cdot \N,
\end{equation}
as shown in more detail in Appendix \ref{appendix}.

\begin{algorithm}
\caption{Split RFD Euler--Maruyama (SEM) scheme \label{alg:sem}}
\begin{enumerate}
	
 \item Generate random forces and torques for all bodies $p$,
 \[
\V{W}_p^{\text{FT}} = k_{B} T \begin{bmatrix}
          \frac{1}{L}_p\V{W}_p^{\text{f}} \\
          \V{W}_p^{\tau}.
\end{bmatrix}
\]

\item Solve the constrained Stokes system,\label{step:solve1}
\begin{equation*}
  \begin{bmatrix}
         -\eta \Lm & \Gm & -\S^{n} & 0 \\
         -\Dm & 0 & 0 & 0 \\
         -\J^{n} & 0 & 0 & -\K^n \\
          0 & 0 & -(\K^n)^{T} & 0
         \end{bmatrix}
         \begin{bmatrix}
         \V{v}^{RFD} \\
         \pi^{RFD} \\
          {\V{\lambda}}^{RFD} \\
          \V{U}^{RFD}
	 \end{bmatrix} = 
	 \begin{bmatrix}
          0 \\
          0 \\
          0 \\
          \V{W}^{FT}
	 \end{bmatrix}.
 \end{equation*}
 
 \item Generate randomly-displaced configurations for all bodies $p$,
 \begin{align*}
 \V{q}_p^{\pm} &= \V{q}_p^{n} \pm \frac{\delta}{2} L_p \V{W}_p^{\text{f}} \\
 \V{\theta}_p^{\pm} &= \text{Rotate}\left(\V{\theta}_p^{n},\pm \frac{\delta}{2} \V{W}_p^{\tau} \right).
\end{align*}

 \item Solve the unconstrained Stokes system \label{alg:uncon}
 \begin{equation*}
  \begin{bmatrix}
          -\eta \Lm & \Gm \\
         -\Dm & 0 
         \end{bmatrix}
         \begin{bmatrix}
          \V{v}^{\#} \\
          \pi^{\#}
	 \end{bmatrix} = 
	 \begin{bmatrix}
          \frac{1}{\delta} \left[\S \left(\V{Q}^{+}\right) - \S \left(\V{Q}^{-}\right) \right] {\V{\lambda}}^{RFD} \\
          0
	 \end{bmatrix}.
 \end{equation*}
 
 \item Compute the random finite differences\label{alg:RFDbits}
 \begin{align*}
 \V{D}^{\K^{T}} &= \frac{1}{\delta} \left[ \K^{T} \left(\V{Q}^{+}\right)  - \K^{T} \left(\V{Q}^{-}\right) \right] {\V{\lambda}}^{RFD} \\
 \V{D}^{\K} &= \frac{1}{\delta} \left[ \K \left(\V{Q}^{+}\right)  - \K \left(\V{Q}^{-}\right) \right] \V{U}^{RFD} \\
 \V{D}^{\J} &= -\frac{1}{\delta} \left[ \J \left(\V{Q}^{+}\right)  - J \left(\V{Q}^{-}\right) \right] \V{v}^{RFD} \\
 \V{D}^{\S} &= \J \V{v}^{\#}.
 \end{align*}
 
 \item Compute the velocities of the rigid bodies by solving the constrained Stokes system\label{step:solve2}
 \begin{equation*}
  \begin{bmatrix}
         -\eta \Lm & \Gm & -\S^{n} & 0 \\
         -\Dm & 0 & 0 & 0 \\
         -\J^{n} & 0 & 0 & -\K^n \\
          0 & 0 & -(\K^n)^{T} & 0
         \end{bmatrix}
         \begin{bmatrix}
         \V{v}^{n} \\
         \pi^{n} \\
          {\V{\lambda}}^{n} \\
          \V{U}^{n}
	 \end{bmatrix} = 
	 \begin{bmatrix}
          \sqrt{\frac{2 \eta k_B T}{\D{t}\D{V}}} \DTm \V{W} \\
          0 \\
          \V{D}^{\K} - \V{D}^{\J} - \V{D}^{\S} \\
          -\V{F}^{n} + \V{D}^{\K^T}
	 \end{bmatrix}.
 \end{equation*}
 
\item Update the positions and orientations of all bodies $p$,
\begin{align*}
 \V{q}_p^{n+1} &= \V{q}_p^{n} + \D{t} \V{U}_p^{n} \\
 \V{\theta}_p^{n+1} &= \text{Rotate}\left(\V{\theta}_p^{n}, \D{t} \V{\omega}_p^{n} \right).
\end{align*} 
\end{enumerate}
\end{algorithm}

\section{Numerical Results}\label{numerical}

To numerically investigate the RB-FIB algorithm, we have implemented it in the IBAMR code \cite{IBAMR}, freely available at \url{https://github.com/IBAMR}, by modifying existing codes developed for deterministic Stokesian suspensions in \cite{RigidMultiblobs}. In particular, we have reused the existing linear solvers in steps \ref{step:solve1} and \ref{step:solve2} of Algorithm \ref{alg:sem}. The scaling and convergence of the numerical linear algebra routines, as well as optimal parameters for the outer and inner Krylov and multigrid iterative solvers are discussed in \cite{RigidMultiblobs}. In all of the following numerical examples (including the Appendix), we use an absolute tolerance proportional to the time step size in the outer FGMRES solver required by steps \ref{step:solve1} and \ref{step:solve2} of the SEM scheme, as well as in the unconstrained GMRES solve required by step \ref{alg:uncon}.
As recommended for the 6-point immersed boundary kernel in \cite{RigidMultiblobs}, we take the spacing of the Lagrangian blobs $s \approx 3 h$, where $h$ is the Eulerian grid spacing. Time steps which generate unphysical configurations (such as a blob overlapping the wall) are rejected and the step repeated. However, these instances are very rare because we use repulsive potentials to prevent particle-particle and particle-wall overlaps, and because we employ the modifications of $\S$ and $\J$ near the boundaries introduced in Appendix D in \cite{RigidIBM}. In all of the following examples, the fluid is water at room temperature $T = 300$ K and viscosity $\eta = 1$ mPa s.

Data for the examples studied in this section was gathered on Northwestern University's QUEST computing cluster. Multiple independent trajectories are run for each case considered in both of the following examples as well as for Appendix \ref{sec:boom}, where we take as many time steps as would complete in a fixed amount of computation time (typically on the order of one week). We accounted for the variable lengths of the runs by using means and variances weighted by the trajectory length when computing relevant statistics. The error bars included in the figures of this section show $95\%$ confidence intervals, computed using the weighted means and variances.

In Appendix \ref{sec:boom} we examine a single colloidal boomerang suspended in a slit channel to demonstrate the first order weak accuracy of the SEM scheme and validate our implementation. This appendix also demonstrates the viability of the RB-FIB method for arbitrary particle shapes and shows the importance of numerically capturing the stochastic drift term in \eqref{LangevinN}.

In Sections \ref{twoSphere} and \ref{lattice} we numerically investigate how tight physical confinement affects hydrodynamic coupling between particles. In section \ref{twoSphere} we use a simple example of two spheres confined in a cuboid in order to establish the spatial resolution required for the RB-FIB method to capture dynamic statistics with sufficient accuracy. In section \ref{lattice} we numerically investigate a variant of an experiment reported in \cite{Bohlein2011} that measured the friction forces and resulting wave patterns in a hydrodynamically driven colloidal monolayer. In our setup the monolayer is confined in a narrow slit channel and is also confined in the lateral directions by walls, mimicking an experiment performed in a microfluidic channel. Using the RB-FIB method we find some novel behavior in the propagation of the waves through the mololayer.

\subsection{Two Spheres in a Tight Cavity}\label{twoSphere}

In this section we investigate how spatial resolution effects the accuracy of the RB-FIB method. We simulate two neutrally buoyant spheres of hydrodynamic radius $R_{h} = 0.656 \ \mu$m, tightly confined in a $3.478 \times 1.739 \times 1.739 \ \mu \text{m}^3$ rectangular box. These physical dimensions ensure that the spheres are almost always in near contact with each other and/or a physical boundary, allowing us to highlight that the RB-FIB method can tackle problems in which confinement plays an important role in the dynamics.

To reduce the severe time step size restriction required to ensure that there are no particle-particle or particle-wall overlaps, we introduce a soft repulsive potential between the two particles as well as between the particles and the wall of the form 
\begin{equation}\label{Usoft}
\Phi(r) = \Phi_0 
\begin{cases}
1 + \frac{d-r}{b} & r < d \\
\text{exp}\left( \frac{d-r}{b} \right) & r \geq d
\end{cases}.
\end{equation}
For the interparticle repulsion, $d = 2 R_H$ and $r$ is the distance between the particle centers. For the repulsion between a particle and a wall, $d = R_H$ and $r$ is the distance between a particle center and a wall. We take $b = 0.1 R_H$ and $\Phi_0 = 4 k_B T$ for both the particle-particle and particle-wall potentials as in \cite{MagneticRollers}, as this choice ensures that the time scale associated with the steric repulsion isn't much smaller than the diffusive time scale, while also maintaining a low probability of unphysical configurations. We use a dimensionless time step size $\Delta \tau = \frac{k_B T}{6 \pi \eta R_{h}^3} \D{t} = 0.0044$ as this was found to be small enough to ensure that the temporal integration errors are smaller than the statistical errors.

\begin{figure}
	\centering
	\includegraphics[width=\textwidth]{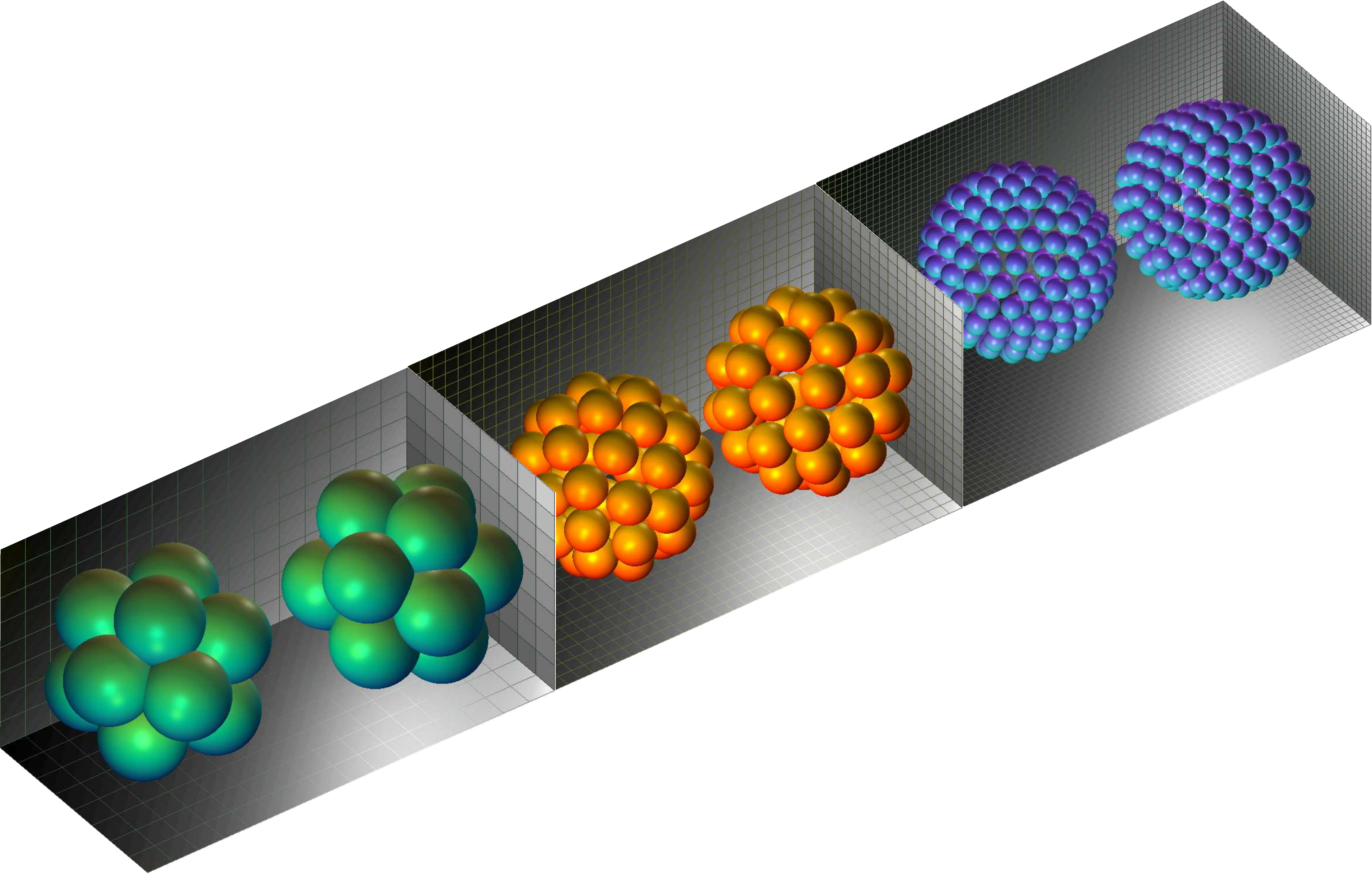}
	\caption{Comparison of different spatial resolutions for two tightly confined spherical colloids where each case is shown in an adjacent cell. Blobs are depicted with their appropriate hydrodynamic radii and the grid spacing of the Eulerian mesh can be seen on the walls of each cell. From left to right, the grid spacing is $h = 0.0543 \times (4,2,1) \mu$m and the spheres are discretized using $12,42,162$ blobs.}
	\label{fig:TwoSphere}
\end{figure}

We discretize the domain using a grid spacing $h = 0.0543 \times (1,2,4) \mu$m and discretize the particles with $162, 42, 12$ approximately equally spaced blobs respectively, as shown in figure \ref{fig:TwoSphere}. The geometric radius of the particles is determined according to table I in \cite{RigidMultiblobs} in order to maintain a constant hydrodynamic radius $R_{h} = 0.656 \mu$m as the resolution is refined. The physical domain is taken to be of dimensions $2L \times L \times L$ where $L = 8 \times 0.0543 \mu$m. 

\begin{figure}
	\centering
	\includegraphics[width=\textwidth]{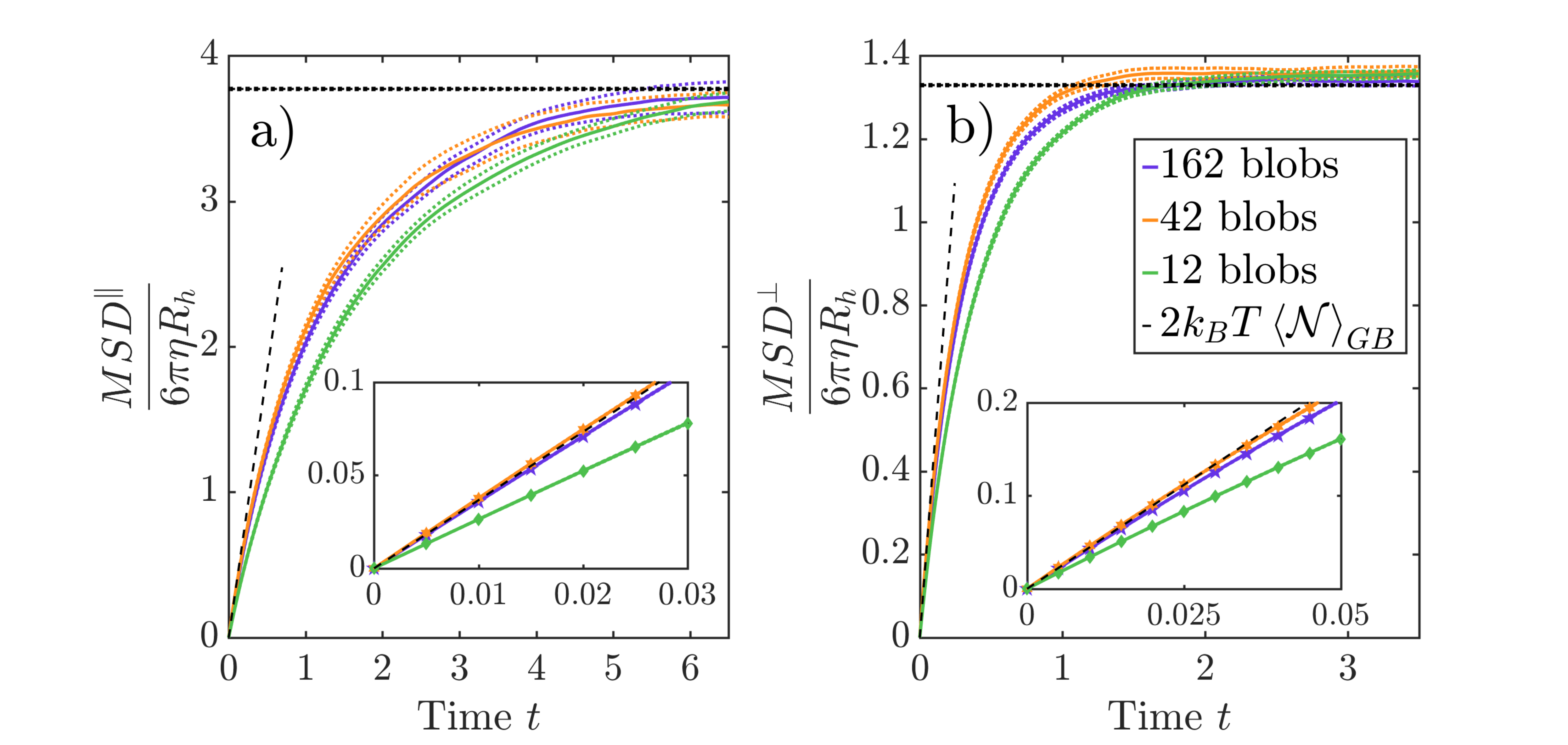}
	\caption{Components of the self MSD ($\msd_{11} = \msd_{22}$) normalized by the free space mobility for different spatial resolutions (see legend). Dashed lines indicate $95\%$ confidence intervals. Also shown as a dashed black line is the normalized short time diffusion component computed according to \eqref{SErel}. The dotted black line shows the asymptotic long-time value as computed using \eqref{asymRef}. The insets in panels (a),(b) show the respective components of the MSD at short times. (a) MSD in the $\parallel$ (long) direction for the two spheres.  (b) MSD in the $\perp$ ($y,z$) directions.}
	\label{fig:TwoSphereMSD}
\end{figure}

In appendix \ref{sec:boom} we study the temporal integration errors by examining the marginals of the equilibrium (static) Gibbs--Boltzmann (GB) distribution for a boomerang confined in a slit channel; we performed similar tests for the two-sphere example studied here and found negligible errors in the static distributions for all resolutions. This is expected for any equilibrium average for sufficiently small time step sizes because only dynamic quantities are affected by the resolution of the hydrodynamics. Here we investigate dynamic statistics in the form of the equilibrium mean squared displacement (MSD) block matrix with tensor blocks
\begin{equation}
 \msd_{pr}\left(t\right) = \av{\left( \V{q}_p (t) - \V{q}_p (0) \right) \left( \V{q}_r (t) - \V{q}_r (0) \right)^{T} },
\end{equation}
where the average is taken over equilibrium trajectories and the subscript $p,r = 1,2$ denotes the particle. 
To compute the components of the MSD we use the SEM scheme to simulate several independent equilibrium trajectories for each of the three resolutions.
In Fig. \ref{fig:TwoSphereMSD} we compare the components of the MSD for different spatial resolutions. Because of the symmetry in the problem, $\msd_{11} = \msd_{22}$ and $\msd_{pp}^{yy} = \msd_{pp}^{zz} \equiv \msd_{pp}^{\perp}$, so we show the average of the equivalent components of the MSD tensor.

Because the particles are completely confined, every component of the translational MSD approaches an asymptote
\begin{equation} \label{asymRef}
\lim_{t \rightarrow \infty} \msd_{pr}\left(t\right) = \av{(\V{z}^p_1 - \V{z}^p_2) (\V{z}^r_1 - \V{z}^r_2)^{T}}, 
\end{equation}
where $\V{z}^{p}_{1}, \V{z}^{p}_{2}$ are two independent samples from the equilibrium distribution of particle $p$, generated from an MCMC method. The appropriate asymptotes are plotted in Figs. \ref{fig:TwoSphereMSD}(a),(b) and we can see that the correct asymptotic MSD is approached using all three resolutions considered. This is again expected because the asymptotic MSD is controlled by the Gibbs-Boltzmann equilibrium distribution and not by the (hydro)dynamics.

The Stokes--Einstein relation gives the short time diffusion tensor
\begin{equation} \label{SErel}
 \V{D}_{pr} = \lim_{t \rightarrow 0} \frac{\msd_{pr} \left(t\right)}{t} = 2 k_{B} T \av{\N_{pr}^{(tt)}}_{GB},
\end{equation}
where the superscript in $\av{\N_{pr}^{(tt)}}_{GB}$ refers to the translation--translation block of the body mobility matrix  $\N$ and $\av{\cdot}_{GB}$ denotes an average with respect to the Gibbs--Boltzmann (GB) distribution. To compute $\av{\N}_{GB}$ we use $642$ blobs to discretize each sphere ($h = 0.5 \times 0.0543 \mu$m) and compute a sample mean of $\N$ over equilibrium configurations sampled using a Markov Chain Monte Carlo (MCMC) method \footnote{We generate enough samples to measure each component of $\av{\N_{pr}^{(tt)}}_{GB}$ to within $1 \%$ statistical error with $95\%$ confidence.}.
The insets of figure \ref{fig:TwoSphereMSD} show that the short-time Stokes--Einstein relation \eqref{SErel} is accurately maintained for both of the finer resolutions but not for the coarsest resolution. Theoretical results are unavailable for the self-diffusion coefficient at intermediate times. We see in Figs. \ref{fig:TwoSphereMSD}(a),(b) close agreement between the two higher resolutions ($162$ and $42$ blobs), but with some visible deviations for the lowest resolution case ($12$ blobs), indicating insufficient resolution.  

\begin{figure}
 \centering
 \includegraphics[width=\textwidth]{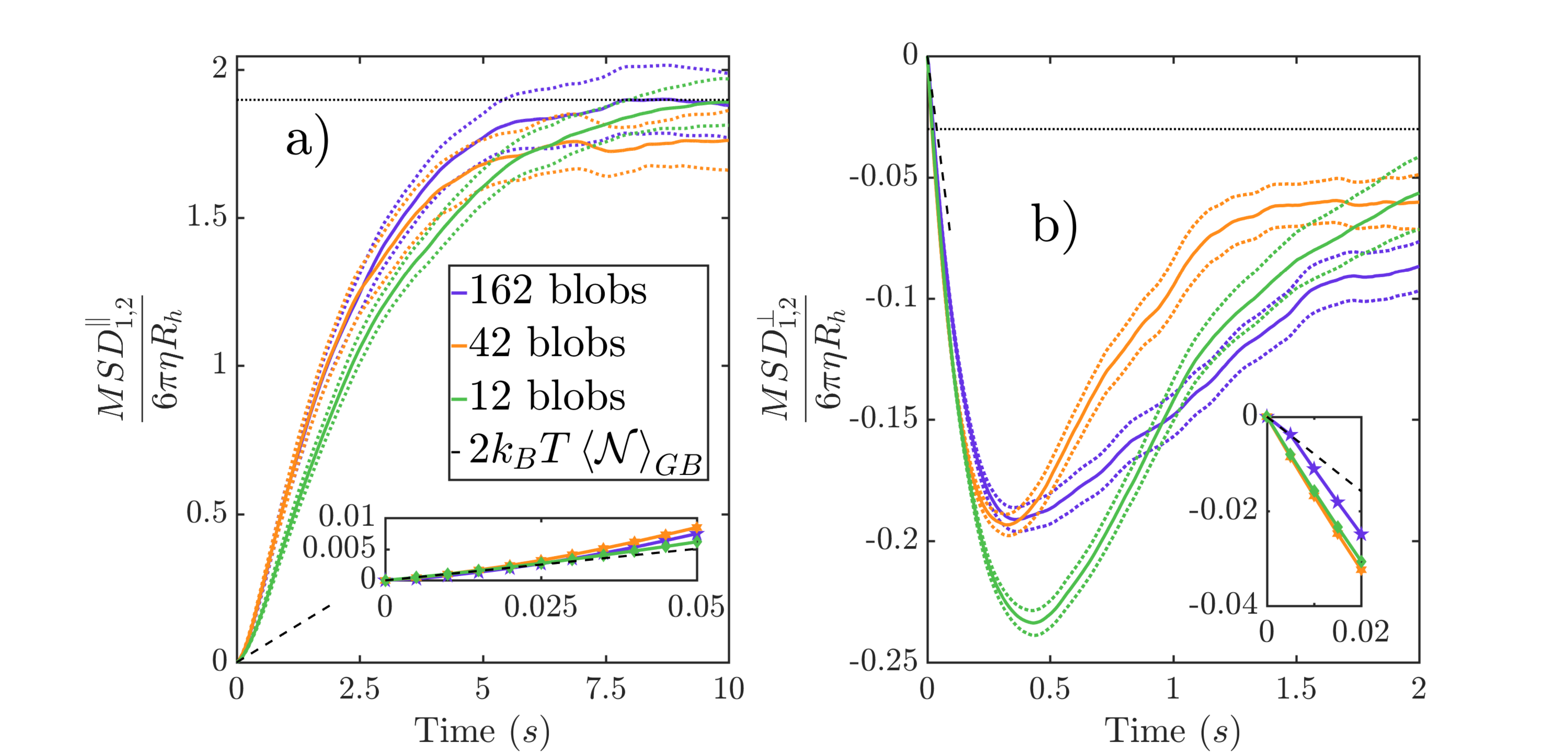}
 \caption{Components of the cross MSD ($\msd_{12} = \msd_{21}$) normalized by the free space mobility for different spatial resolutions (see legend). Dashed lines indicate $95\%$ confidence intervals. Also shown as a dashed black line is the normalized short time diffusion component computed according to \eqref{SErel}. The dotted black line shows the asymptotic long-time value as computed using \eqref{asymRef}. The insets zoom in on the short-time behavior. (a) MSD in the $\parallel$ (long) direction. (b) MSD in the $\perp$ ($y,z$) directions.}
 \label{fig:TwoSphereMSDcross}
\end{figure}

The authors of \cite{TwoSphereCouple} found analytically and experimentally that two nearby colloidal spheres were strongly hydrodynamically coupled but the coupling weakens significantly near a confining boundary. In our example we see a competition of influence: the two spheres are always close to each other and hence their motion should be coupled, but they are also always close to a physical boundary which would decouple their motion. To get a clearer picture of this competition, Fig. \ref{fig:TwoSphereMSDcross} shows the non-vanishing components (the $xx$ or $\parallel$ and $yy,zz$ or $\perp$ components) of $\msd_{1,2}$. The inset of Fig. \ref{fig:TwoSphereMSDcross}(a) shows that the short time diffusive motion in the $\parallel$ direction between the two spheres (given by \eqref{SErel}) is very weakly coupled. However, the full panel of Fig. \ref{fig:TwoSphereMSDcross}(a) shows a much stronger coupling in the parallel motion of the spheres for longer times. Therefore, the influence of another nearby particle eventually dominates over the influence of the walls, which initially decouples the particles' motion. In Fig. \ref{fig:TwoSphereMSDcross}(b) we see that the short time diffusion in the $\perp$ ($y,z$) directions between the spheres is somewhat strongly anti-correlated. After $t \approx 4$s the behavior of the MSD inflects and decays to nearly zero for larger times and the perpendicular motion of the spheres effectively decouples due to the confinement. The results in Fig. \ref{fig:TwoSphereMSDcross} indicate that 12 blobs per sphere is not sufficient to accurately predict the time correlation functions for more than one particle, and at least 42 blobs per sphere are required for particles this close to each other and the walls.

\subsection{Friction in a Colloidal Monolayer}\label{lattice}

In \cite{Bohlein2011}, the authors performed an experiment in which a colloidal monolayer was hydrodynamically driven across a bottom wall on which a substrate potential was generated by optical traps. This potential mimics the effect of corrugation of the wall, which is a key contributor to the effective friction with the wall. In \cite{Bohlein2011} the system was kept quasi--two--dimensional by forcing the monolayer to remain near the wall using a vertically incident laser to form a confining potential that dramatically reduced out of plane motion. The colloids used in the experiment were negatively charged polystyrene spheres suspended in water. Due to their negative charge, the particles spontaneously formed a stable 2D triangular crystal \cite{YukawaLattice20KT}. The corrugation potential used in the experiments reported in \cite{Bohlein2011} took the form of a periodic lattice with 3-fold symmetry around its minima. The colloidal crystal and the corrugation potential are called commensurate if the lattice constants agree and incommensurate otherwise; both cases were considered in \cite{Bohlein2011}. 

In the experiments reported in \cite{Bohlein2011}, the sample cell was translated to generate a flow field and hence a fluid drag force on the colloids. This lateral drag force on the crystal served as a control parameter and, under commensurate conditions, the authors observed a critical translation velocity (and hence lateral force) at which the colloidal crystal became unpinned from the corrugation potential. This critical value represents the transition of the crystal from static friction to kinetic as it becomes unpinned. Just above this critical velocity, they observed localized density variations in the colloidal crystal taking the form of traveling kink solitons.

\begin{figure}
	\centering
	\includegraphics[width=\textwidth]{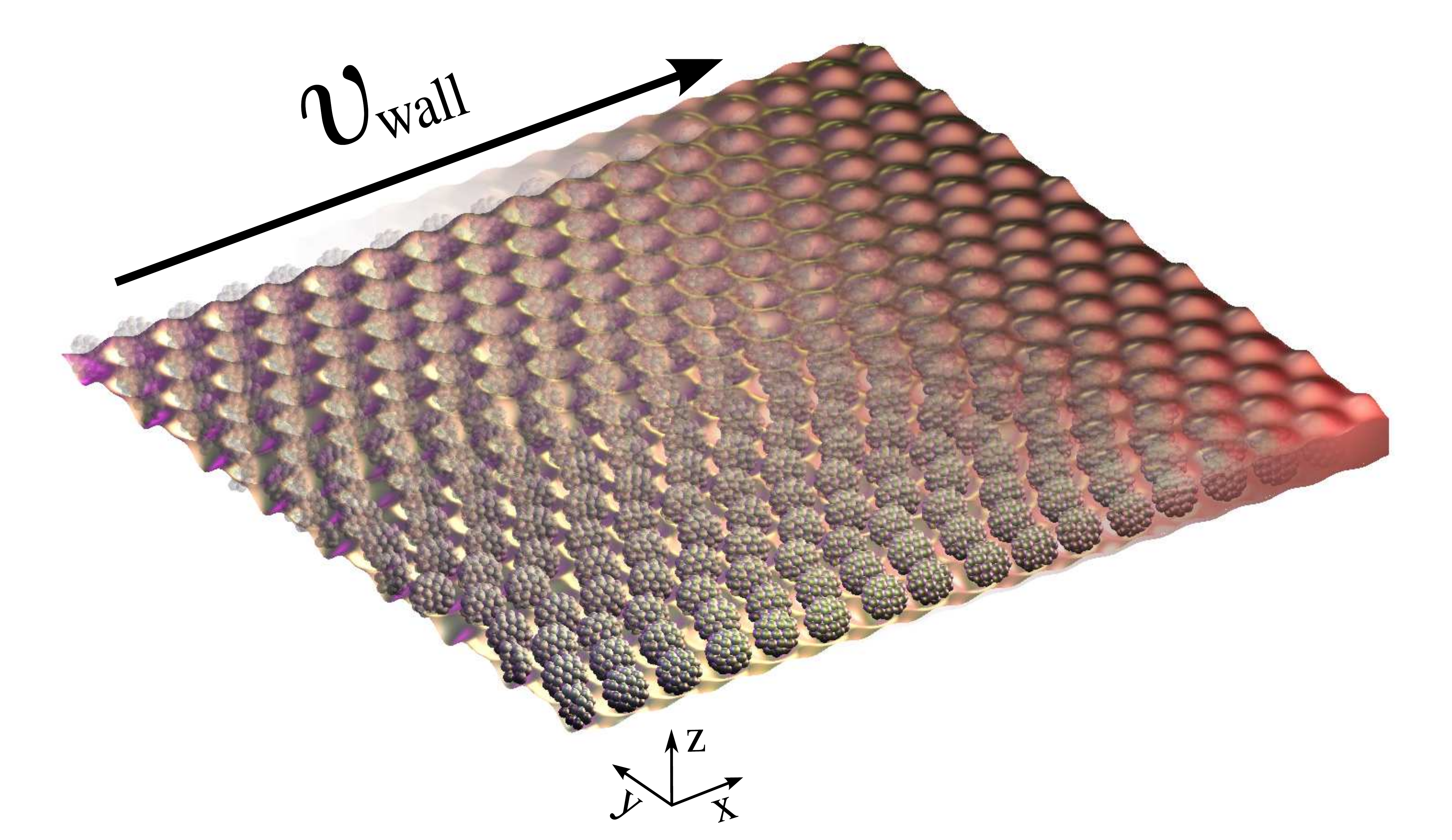}
	\caption{A diagram of a typical simulation configuration that we consider in this section. Quasi--two dimensionality is achieved through screened Coulombic interactions between the particles (represented here using their 42--blob discretizations) and the two walls in the $z$-direction. Walls at the boundaries in the $y$-direction are also shown. The physical boundaries themselves are represented as corrugated sheets (transparency is added in the $xy$-direction for visual clarity) to represent the periodic substrate potential used in the simulations (note however that this is simply for visualization and the physical boundaries are indeed flat). A drag force is applied to the colloidal monolayer by prescribing a wall velocity $v_{\text{wall}}$ in the positive $x$ direction (flow direction shown as a black arrow). Boundaries of the domain in the $x$-direction are taken to be periodic to allow for continuous movement of the monolayer.} 
	\label{fig:lattice}
\end{figure}

In this section, we will use the RB-FIB method to numerically investigate a novel variant of this experiment wherein the quasi--two--dimensionality of the problem (essential for the existence of a stable 2D monolayer) is achieved by confining the monolayer in a slit channel between two walls, see illustration in figure \ref{fig:lattice}. This type of confinement is easy to realize in experiments using a microchannel. Additionally, this example serves to demonstrate an important feature of the RB-FIB method: not only can it account for tight confinement, but it can also easily account for external flow fields generated by the imposed boundary conditions.
  
\subsubsection{Physical Parameters}

The electrostatic repulsion between the colloids is accounted for by a pairwise Yukawa potential of the form
\begin{equation} \label{YukawaPot}
 U_{\text{pair}}(r) = U_0 \frac{\exp\left(\frac{D_H - r}{\lambda_D} \right)}{r/  D_H},
\end{equation}
where $U_0$ is the repulsion strength, $\lambda_D$ is the Debye screening length, and $r$ is the distance between particle centers. We set the screening length $\lambda_D=0.16 \mu$m to the value reported in \cite{Bohlein2011}. The repulsion strength $U_0$ is a free parameter which must be chosen to ensure that a stable colloidal crystal is formed in the absence of a trapping potential, and that this crystal remains stable in the presence of the thermal fluctuations and background flow. In agreement with \cite{YukawaLattice20KT,YukawaLattice80KT}, we find that choosing $20 k_{B} T \leq U_0 \leq 80 k_{B} T$ is reasonable as the results presented later in this section were not strongly effected by taking $U_0$ to be either of these extremes; we will fix $U_0 = 20 k_{B} T$ henceforth. To reduce the probability of blob--wall overlaps we include a soft repulsive potential between each blob and all of the physical boundaries of the domain. This potential takes the form of \eqref{Usoft}, where $d = a$, $b = 0.1 a$, and $\Phi_0 = 4 k_B T$, and $r$ is the distance between a blob and a wall.\footnote{For spherical particles we could have put a wall-repulsive potential on each sphere's center rather than each blob and avoided small spurious torques on the particles. However this would not generalize easily to arbitrary particle shapes.} We use a dimensionless steric time step size $\Delta \tau = \D{t} \left(\Phi_0/(6 \pi \eta a^2 b)\right)  = 0.138$ which we found to be sufficiently small to make particle--wall overlaps infrequent.

To capture the effects of corrugation, each colloid feels the potential \cite{ColloidSheetFriction}
\begin{equation}
U_{\text{corr}}\left(\V{X}_{2D}\right) = \frac{U_{0}^{\text{corr}}}{2} \left\{ 3 - \cos \left( \left[ \V{k}_1 - \V{k}_2 \right] \cdot \V{X}_{2D} \right) - \cos \left( \V{k}_1 \cdot \V{X}_{2D} \right)- \cos \left(  \V{k}_2 \cdot \V{X}_{2D} \right) \right\},
\end{equation}
where a particle's $y$--symmetrized 2D position is $\V{X}_{2D} = [x, y-L_y / 2]^{T}$.
Here the scaled lattice directions are
\begin{equation}
 \V{k}_1 = \frac{4 \pi}{\sqrt{3} s_{\text{Lattice}}} \begin{bmatrix}
            -\frac{\sqrt{3}}{2} \\
            -\frac{1}{2}
           \end{bmatrix}, \hspace{1cm}
 \V{k}_2 = \frac{4 \pi}{\sqrt{3} s_{\text{Lattice}}} \begin{bmatrix}
            -\frac{\sqrt{3}}{2} \\
            \frac{1}{2}
           \end{bmatrix}.
\end{equation}
Following \cite{Bohlein2011}, we take the average particle separation to be $s_{\text{Lattice}} = 5.7 \mu$m and the particle radius $R_H = 1.95$. In the absence of a substrate potential the spacing of the colloidal crystal is controlled by the number of particles and the domain dimensions. To ensure that the spacing of the substrate potential $s_{\text{Lattice}}$ is commensurate with the spacing in the colloidal crystal, we use 272 particles and take $L_x = 16 s_{\text{Lattice}}$, $L_y = 15 s_{\text{Lattice}}$. The domain is taken to be periodic in the direction of the applied flow field ($x$ direction) and no--slip boundaries are used in every other direction. While the periodicity in the $x$ direction introduces some unphysical artifacts, $L_x$ is large enough to produce kink solitons, which were found to have a support of $\approx 8 s_{\text{Lattice}}$ \cite{Bohlein2011}. The width of the domain in the $z$ direction is taken to be $L_z = 1.28 D_H$, where $D_H = 2 R_H$ is the particle diameter. This is to ensure the fairly strict quasi--two dimensionality required for colloidal crystals to form.

To drive the colloidal monolayer, we move the top, bottom and side walls with velocity $v_{\text{wall}}$ along the $x$ axes, i.e., we impose a fluid velocity $\left(v_{\text{wall}},0,0\right)$ on all walls as a boundary condition for the Stokes equations 
\footnote{In this simple case, the flow created by the wall slip is simple constant plug flow, so one could move/shift the corrugation potential with velocity $\left(-v_{\text{wall}},0,0\right)$ instead of imposing a velocity on the walls. However, in more general situations, e.g., (time-dependent) pressure-driven flow in square channels, it is much simpler and more flexible to let the Stokes solver compute the fluid flow generated by the boundaries.}.
We non--dimensionalize the control parameter $v_{\text{wall}}$ using the work required to move one colloid one lattice site in an unbounded domain,
\begin{equation}
 W_{\text{wall}} = \frac{6 \pi \eta R_H v_{\text{wall}} \cdot s_{\text{Lattice}} }{U_{0}^{corr}}.
\end{equation}
In what follows, we vary $W_{\text{wall}} = \left\{ 1.63, 1.86, 2.1, 2.33, 2.56\right\}$ and investigate the dynamics of the colloidal monolayer shown in figure \ref{fig:lattice}.

\subsubsection{Discretization Parameters}

We discretize the colloids using 42 blobs, as this resolution was found in section \ref{twoSphere} to provide sufficient accuracy for the dynamics. The fluid grid spacing $h=s_{\text{Lattice}}/16$ so that substrate potential aligns with the periodic domain size in the $x$ direction. The blob spacing $s = 0.95$ was chosen so that the particles' hydrodynamic radius would be $R_H = 1.95$ (geometric radius $R_{G} = 1.7380$) and $s \approx 3 h$ as was recommended in \cite{RigidMultiblobs}.

\subsubsection{Numerical Observations of Kinks in a Colloidal Monolayer}

In this section we vary the dimensionless wall velocity $W_{\text{wall}}$ to investigate the critical value at which static friction is broken and the colloidal monolayer begins to slide through the thin 'corrugated' channel via kink solitons. To measure the point at which the colloidal monolayer breaks from the corrugated substrate potential, we simply determine whether any particle displaces more than $s_{\text{Lattice}}$ from its initial position. Partial depinning of the colloidal crystal is observed at $W_{\text{wall}} = 1.86$ wherein only some of the particles are sufficiently displaced. Full depinning is observed for all $W_{\text{wall}} > 1.86$ wherein all of the particles are sufficiently displaced after some amount of time. Brownian motion is crucial to the formation and propagation of kinks. While it is certainly possible to see depinning of the monolayer in the absence of thermal fluctuations, the tiny 'kicks' provided by the fluctuating fluid activate a particle's transition between potential wells. So much so, in fact, that no depinning was observed in complimentary deterministic simulations for all values of $W_{\text{wall}}$ considered.

As noted in \cite{Bohlein2011}, a kink is formed when one particle escapes the potential well it is confined to through a combination of thermal forces as well as hydrodynamic drag from the background flow. Once a particle escapes, it enters the neighboring potential well in the direction of the flow. Once confined to this new well, which also typically has a particle trapped in it, a combination of Yukawa and steric repulsion forces the original occupant of the well into its neighboring well, and the process repeats. We identify as a kink the propagation of particles escaping their well along the direction of flow. 

\begin{figure}
 \centering
 \includegraphics[width=\textwidth]{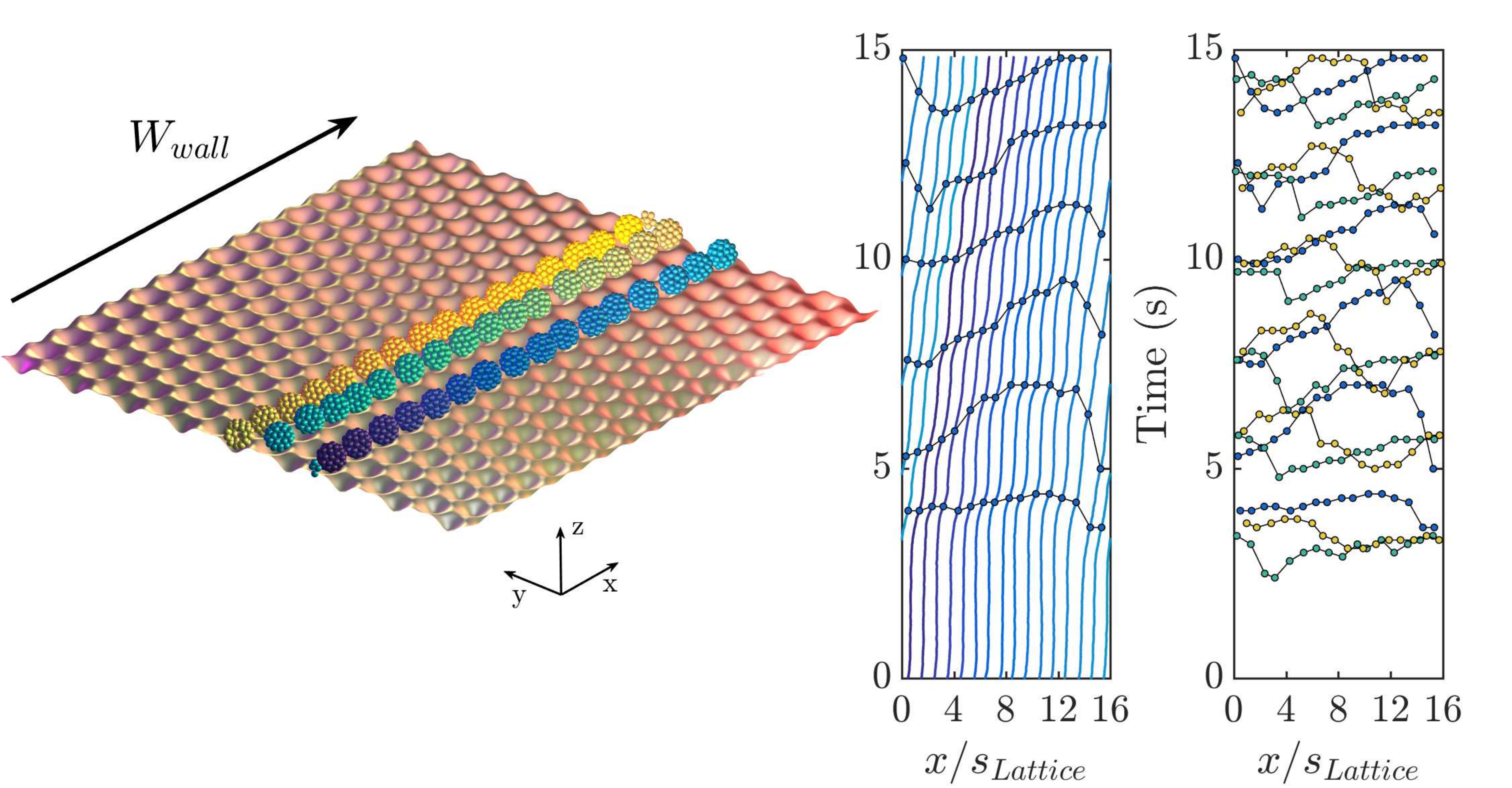}
 \caption{(a) Three rows of colloidal particles in the confined monolayer, highlighted in different color gradients. The rows are at a distance of one (yelow and green), two (green and blue), or three (yellow and blue) rows apart. Note that other particles as well as the top and side walls are not drawn here for visual clarity. (b) Trajectories in the $x$ direction of the rightmost row (blue) of particles shown in panel (a). The color of the trajectories corresponds to the particle color. The black contours which are orthogonal to the particle trajectories are traveling kinks. These contours connect the local maxima of the velocity of each particle, averaged over 50 time steps in order to filter out high frequency fluctuations due to thermal motion. (c) Comparison of the  kink propagation (black lines in panel (b)) for each of the three rows of particles, showing only a small degree of coordination between different rows.}
 \label{fig:kink}
\end{figure}

Figure \ref{fig:kink} shows the propagation of kinks in three nearby rows of particles along the direction of flow; see the SI for a movie. Figure \ref{fig:kink}(b) shows the $x$ position of each particle from the `blue' row  over time. The transverse black contour lines show local extrema in the velocities of the particles (averaged over 50 time steps to filter out the Brownian velocities), which are seen to correspond to the times at which a particle jumps to another lattice site. The bends in the black velocity contours show a finite speed of propagation for the kink and the `S' shaped profile is due to the periodicity in the $x$ direction. Figure \ref{fig:kink}(c) shows the contours of the maximum velocity for the particle rows highlighted in panel (a). The prevailing `S' shape in all of the contours in panel (c) demonstrates that there are propagating kinks in all of the rows of particles in the monolayer. However, it is difficult to appreciate the correlations in the motion of the kinks in nearby rows in these results.

\begin{figure}
 \centering
 \subfloat{\includegraphics[width=\textwidth]{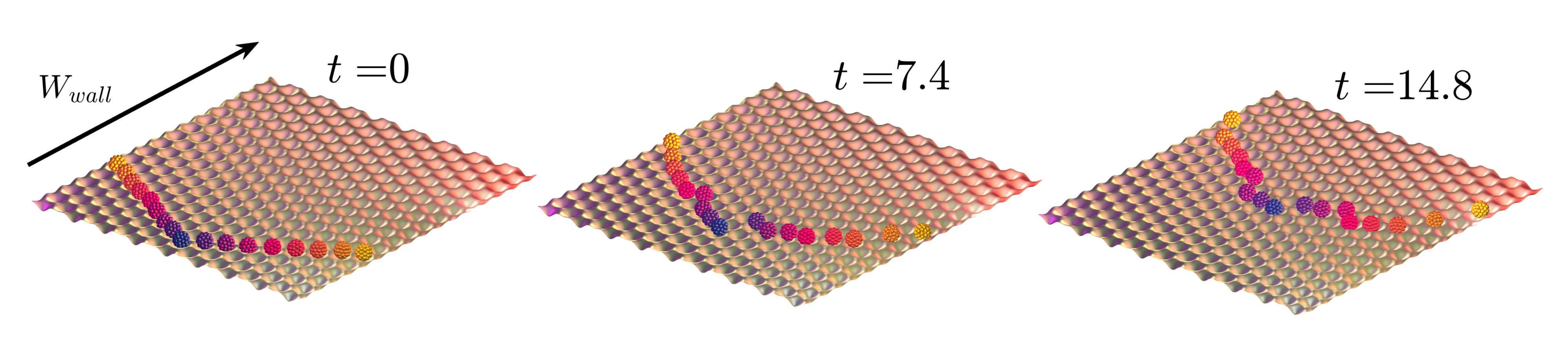}}
 \hfill
 \subfloat{\includegraphics[width=\textwidth]{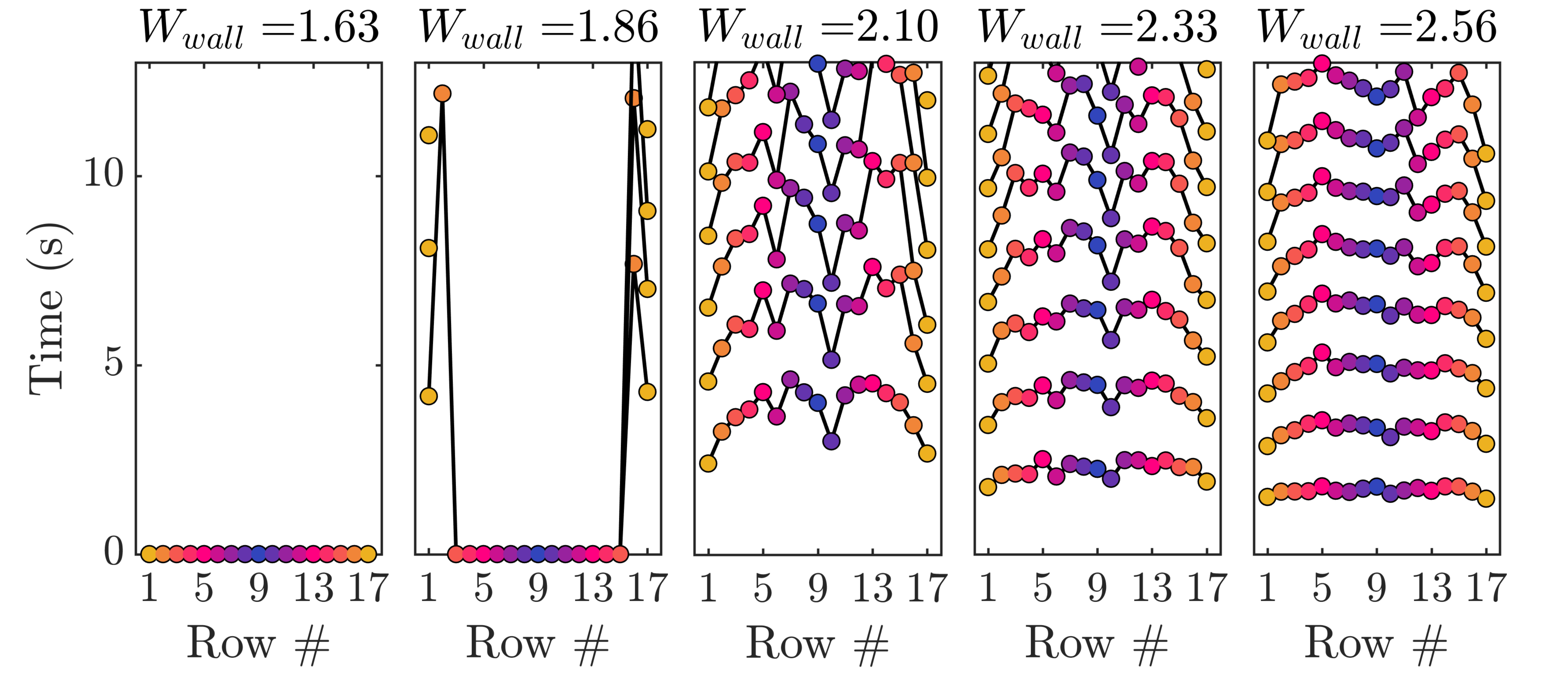}}
 \caption{The top row of panels shows the positions of the colored particles at three different times for $W_{\text{wall}} = 2.10$. The hopping (displacement) times for the colored particles are shown in the bottom row of panels for each particle for different values of $W_{\text{wall}}$. The row number on the $x$ axes corresponds to the $y$ position of the particles, increasing from left to right, and a marker is placed along the time axis whenever the given particle displaces more than a lattice width $s_{\text{Lattice}}$. The first, second, third, etc. hopping times of the particles are connected with a black line. The black lines develop a clear `M' shape in time for each case considered, indicating that the particles near the $y$ boundaries displace first, followed by the particles in the middle.}
 \label{fig:kink_wave}
\end{figure}

In figure \ref{fig:kink_wave} we investigate how kinks influence other kinks in the direction transverse to their motion. The authors of \cite{Bohlein2011} observed that kinks extend also perpendicular to the direction of the force with a small lag in time. Their system however, was large enough to be considered unconfined in the $y$ direction. Here we examine the correlation of kinks between rows of particles, where a row is defined by binning the particles' $y$ coordinates with bin width $2 s_{\text{Lattice}}$. By selecting a representative particle from each row in the monolayer, as seen in the top panels of figure \ref{fig:kink_wave}, we can track the hopping or `displacement' times for each row. That is, we track the time at which the representative particle of a row has been displaced by an integer multiple of $s_{\text{Lattice}}$. The bottom row of panels in Fig. \ref{fig:kink_wave} shows the hopping times for representative particles shown in the top row of panels, where the color of the markers corresponds to the color of the particle. The displacement times of the particles are grouped based on how many lattice sites the particles have displaced. That is, every particle's first displacement time is connected, as is every particle's second, and so on. 

The bottom row of panels in Fig. \ref{fig:kink_wave} shows that for every value of $W_{\text{wall}} > 1.63$, the first rows to be displaced are the rows closest to the walls which bound the domain in the $y$ direction. For $W_{\text{wall}} = 1.87$, the only particles which become displaced are those in the rows nearest these walls and their immediate neighbors. This is likely because of the additional hydrodynamic screening provided by the walls, as we investigated in section \ref{twoSphere}. By analogy with what was observed in \cite{Bohlein2011}, one might expect that the next rows to be displaced are the immediate neighbors of the rows nearest the walls, and then their neighboring rows, and so forth, with the middle rows displacing last. To the contrary, we see that one of the middle rows displaces soon after the displacement of the rows nearest the walls. This is a surprising result and may stem from the lateral ($y$) confinement of the system. A more thorough investigation of the unusual collective dynamics of kinks in this system is deferred to future work.

\section{Conclusion} \label{sec:conc}

In this work, we described the Rigid Body Fluctuating Immersed Boundary (RB-FIB)  method to simulate the Brownian dynamics of arbitrarily shaped rigid particles in fully confined domains. The fluctuating solvent was treated explicitly, allowing for well-known staggered finite difference schemes to be used and general boundary conditions to be applied on the boundaries of the computational domain. We designed an efficient Split--Euler--Maruyama temporal integrator that uses a nontrivial combination of random finite differences to capture the stochastic drift appearing in the overdamped Langevin equation. We implemented this method in the IBAMR software infrastructure, which is freely available at \url{https://github.com/IBAMR}. 

We studied the dynamical correlation functions of two tightly confined spheres in close proximity to each other and physical walls, and examined the effect of spatial resolution on the accuracy of both the short- and long-time equilibrium mean-square displacement.
We also used the RB-FIB method to model a  quasi--2D colloidal crystal confined in a narrow slit channel. The layer was hydrodynamically driven across a commensurate periodic substrate potential mimicking the effect of a corrugated wall. We observed partial and full depinning of the colloidal monolayer from the substrate potential above a certain wall speed, consistent with the transition from static to kinetic friction observed in \cite{Bohlein2011}.  Further, we observed the propagation of kink solitons parallel to the direction of flow. These kinks extended along the colloidal monolayer in the direction transverse to the flow. We observed a curious `M' pattern in the particle displacements across the domain wherein particles nearest the boundaries of the domain \emph{and} particles in the middle of the domain are the first to be displaced.

The SEM scheme presented here is based on the Euler--Maruyama method and as such is first order weakly accurate (as shown both numerically and theoretically in Appendix \ref{sec:boom} and \ref{appendix}), as well as only first order deterministically accurate. Higher order deterministic accuracy has been shown to be very beneficial in designing temporal integrators for Brownian dynamics of rigid particles in half-space domains \cite{BrownianMultiblobSuspensions}. The SEM scheme can be extended to deterministically second-order accurate Adams--Bashforth, trapezoidal and midpoint variants, but whether or not the additional computational cost incurred by these methods is justified should be investigated in future work. 

The RB-FIB method we present here has notable advantages over other methods. First, it can handle a variety of combinations of boundary conditions in different directions seamlessly. Second, it can handle colloidal particles of complex shapes with varying levels of spatial resolution (fidelity), i.e., with controllable accuracy. Third, the method scales linearly in the number of particles and fluid grid cells. Fourth, the explicit solvent approach also facilitates coupling to additional physics, including elastic bodies handled using the immersed boundary method, non-Newtonian or multicomponent solvents, and electrohydrodynamics.
However, there are also some disadvantages compared to other methods. First, because the method uses an explicit representation of the fluid domain, infinite domains cannot be considered in the present formulation. Second, because the method scales linearly in the number of \emph{fluid} and particle degrees of freedom, phenomena involving only a few particles and large fluid domains are particularly inefficient to simulate using the RB-FIB method. Third, lubrication forces between nearly touching particles are not accurately handled for realistic number of blobs per particle, and the method becomes less efficient (in terms of both memory use and computing time) when there are many blobs per particle. Some of these shortcomings can be addressed by using explicit Green's functions \cite{BrownianMultiblobSuspensions}, or specializing to spherical particles \cite{Galerkin_Wall_Spheres,AutophoreticSpheres_Adhikari,BrownianDynamics_OrderNlogN}.

One major advantage of the RB-FIB method not exploited in this work is it's ability to include fluid body forces in the momentum equation. This allows for the inclusion of a range of multiphysics phenomena whose coupling with the fluid is via a fluid body force. An example are electrohydrodynamic phenomena for which the body force is the divergence of the Maxwell stress tensor \cite{Bhalla2014FullyRI}. Since electrostatic fields are non--dissipative, no additional effort is needed to be taken to account for thermal fluctuations. However, charge separation creates a thin Debye layer near solid surfaces which is difficult to resolve numerically, but can be approximated using asymptotics \cite{ElectroneutralAsymptotics_Yariv}. The ability of the RB-FIB method scheme to prescribe active slip velocities on the surfaces of the particles \emph{and} the boundaries can be used to account for electroosmotic slip flows without resolving the Debye layers.

\begin{acknowledgments}
This work was supported by the MRSEC Program of the National Science Foundation under Award Number DMR-1420073.
This work was also partially supported by the National Science Foundation under collaborative
award DMS-1418706 and by DMS-1418672. We thank Northwestern University's Quest high performance computing service for the resources used to perform the simulations in this work. Brennan Sprinkle and Aleksandar Donev were supported in part by the Research Training Group in Modeling and Simulation funded by the National Science Foundation via grant RTG/DMS-1646339. Amneet Bhalla acknowledges research support provided by the San Diego State University. 
\end{acknowledgments}

\appendix

\section{Temporal Accuracy for a Confined Colloidal Boomerang} \label{sec:boom}

\begin{figure}
	\centering
	\includegraphics[width=\textwidth]{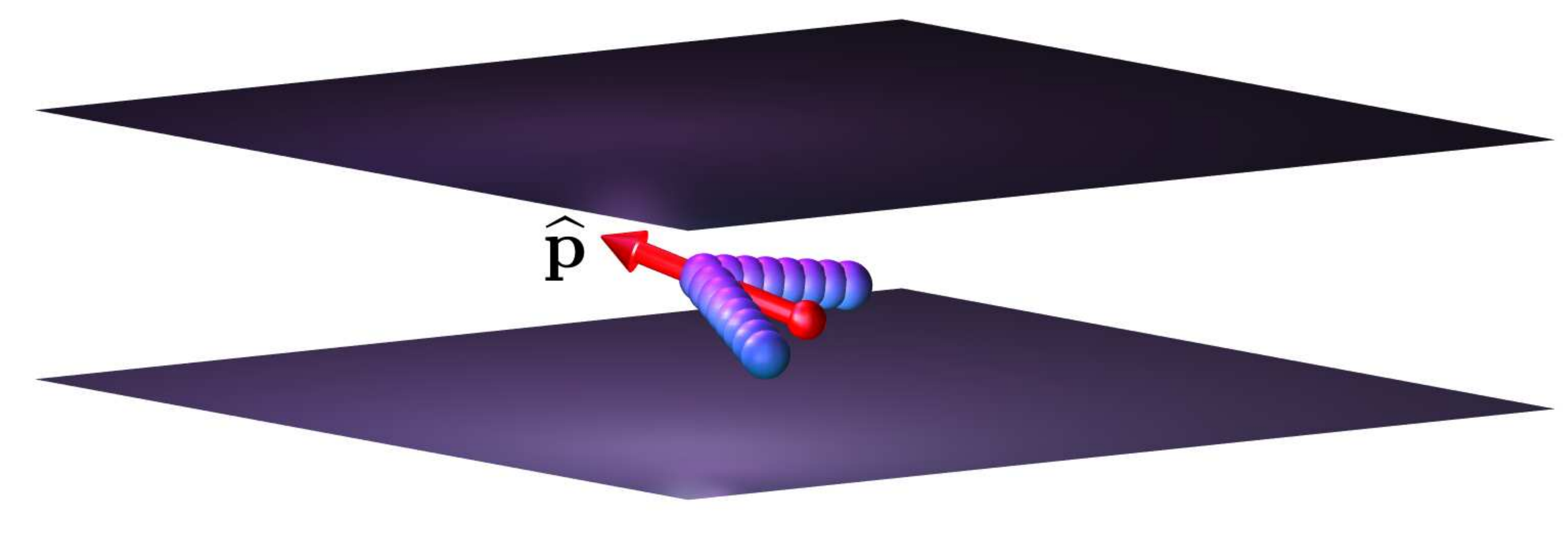}
	\caption{A typical configuration of a colloidal boomerang confined between two parallel walls. The vector, $\hat{\V{p}}$, tracking the boomerang's cross--point orientation is also shown as a red arrow (not to scale).}
	\label{fig:boomcartoon}
\end{figure}

To investigate the weak accuracy of the RB-FIB method scheme, we consider a single colloidal `boomerang' suspended in a periodic slit channel, as depicted in figure \ref{fig:boomcartoon}. In addition to examining the temporal accuracy of the SEM scheme, this example demonstrates the ability of the method to handle arbitrary particle shapes as well as fully confined physical domains.  

The fluid is water at room temperature $T = 300$ K and viscosity $\eta = 1$ mPa s. We model the boomerang using 15 equally spaced blobs with 7 of them forming each of the $2.1 \mu$m arms of the `L' shaped body. We take the blob radius $a$ to be related to the inter--particle spacing $s = 0.3 \mu$m as $s = 0.925 a$, as was recommended in \cite{RigidMultiblobs}. As we employ the 6--point delta kernel \cite{New6ptKernel} here, we have that $a = 1.47 h$ and hence the Eulerian mesh width $h = 0.2208 \mu$m.

The physical domain has extent $64 h \times 64 h \times 16 h$ with no slip conditions on the top and bottom boundaries and periodic conditions in the $x,y$ directions. These dimensions were chosen to have some similar qualitative features to the quasi-2D confinement of colloidal boomerangs in the experiments reported in \cite{BoomerangDiffusion,AsymmetricBoomerangs}; the boomerang \emph{barely} has enough room to rotate fully in any direction if it is centered between the walls. We include a soft repulsive potential between the blobs and the wall of the form \eqref{Usoft} but with $d = a$ and $r$ interpreted as the distance between the blobs and the wall. We use $b = 0.1 R_H$ and $\Phi_0 = 4 k_B T$ as in \cite{MagneticRollers}, as this choice ensures that the steric time scale associated with the potential isn't much smaller than the diffusive time scale, while also maintaining a low probability of blob-wall overlaps. As in \cite{BrownianMultiblobSuspensions}, we give each blob a buoyant mass of $m_e = 1.57 \times 10^{-11}$mg and thus apply a gravitation force $-m_e g \hat{z}$ on each of the 15 blobs, where $g$ is the Earth's gravitational acceleration. 

\begin{figure}
 \centering
 \includegraphics[width=\textwidth]{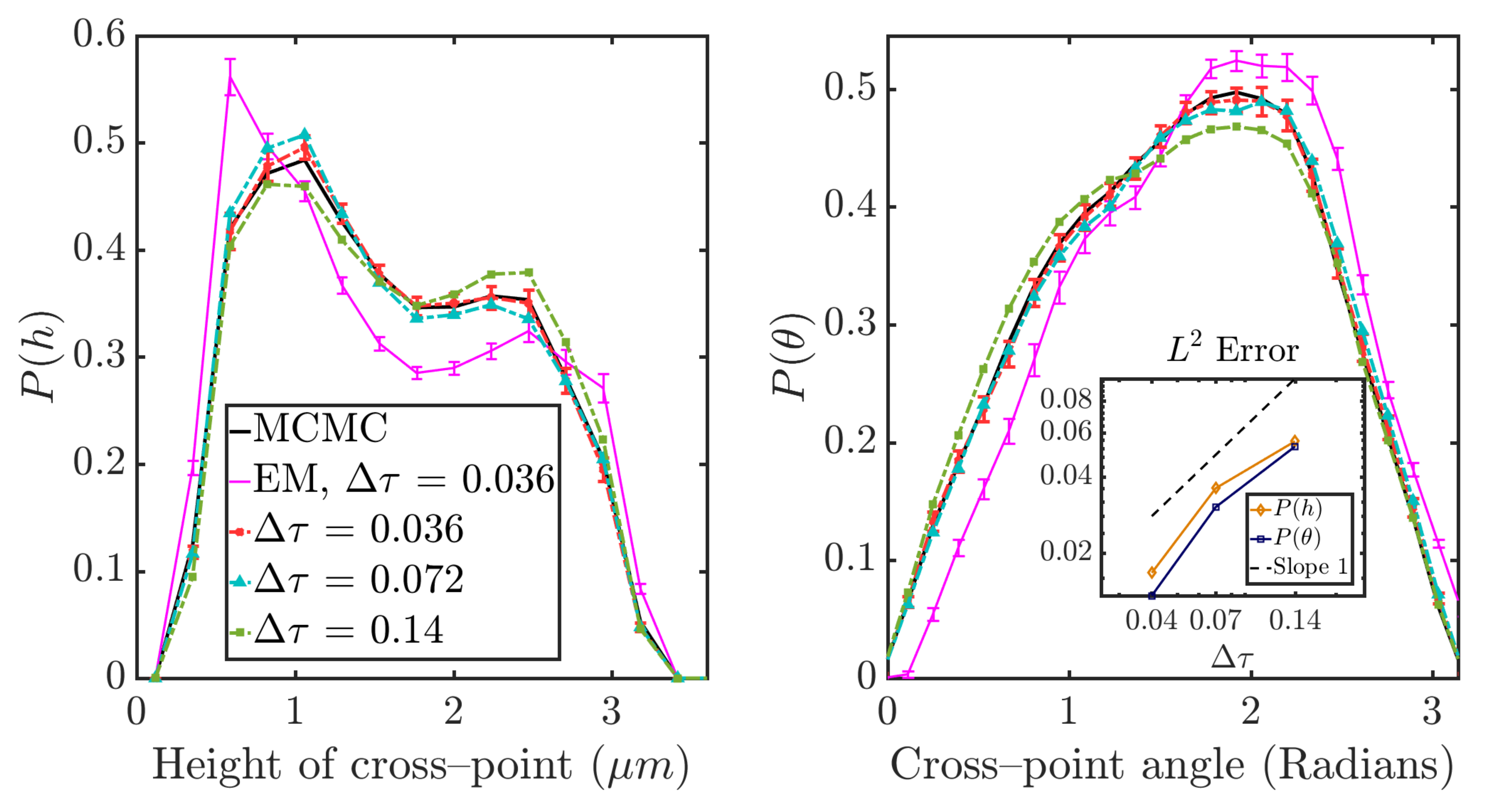}
 \caption{(a) Probability distributions of the height of the cross point of the boomerang for several time step sizes; error bars (only shown for $\Delta \tau = 0.036$) denote $95\%$ confidence intervals. The black curve was generated using direct samples from the equilibrium GB distribution using an MCMC method. The pink curve shows the biased distribution generated by setting the RFD terms in step \ref{alg:RFDbits} of Algorithm \ref{alg:sem} to zero, thereby not producing any stochastic drift (that is, using a naive Euler--Maruyama scheme). (b) Probability distributions of the angle of the boomerang cross point, measured with respect to the positive $z$--axis. Inset: Log--log plot of the $L^2$ error in the distributions of the cross point height and angle, consistent with first-order weak accuracy.}
 \label{fig:boom}
\end{figure}

Figure \ref{fig:boom}(a) shows the probability distribution of the height of the boomerangs cross--point (the blob at the bend in the `L' shape) computed using the SEM scheme with 3 different non--dimensionalized time step sizes, $\Delta \tau = \D{t}\,(k_B T)/(6 \pi \eta a^3) = \left\{ 0.14, 0.072, 0.036\right\}$. The largest time step size considered was chosen to be just under the empirically measured stability limit of $\Delta \tau \approx 0.22$. The black curve represents the correct marginal distribution which we obtain from the equilibrium Gibbs--Boltzmann distribution using an MCMC method. Comparing the marginal distributions generated by the SEM scheme to the correct distribution, we see clear improvement as the time step size is decreased. The inset of figure \ref{fig:boom}(b) shows that this decrease in the error is consistent with the first order weak accuracy proven in Appendix \ref{appendix}. To underscore the importance of accounting for the stochastic drift term in \eqref{LangevinN}, figure \ref{fig:boom} also shows the biased equilibrium distribution obtained by using the SEM scheme without any RFD terms to account for the drift. We can see that there is a marked error in the biased distribution emphasizing the importance of the stochastic drift.

The cross--point height distribution shown in figure \ref{fig:boom}(a) is clearly bimodal. The peaks arise because of the shape of the particle and how it interacts with the confinement of the parallel walls. The asymmetry in the two peaks of distribution is partly due to the gravitational force. To interrogate this further, we track a unit vector $\hat{\V{p}}$ (shown in figure \ref{fig:boomcartoon}) that is in the plane of the boomerang and intersects the boomerang's cross--point at $45^{\circ}$. Figure \ref{fig:boom}(b) shows the distribution of the angle $\theta$ that $\hat{\V{p}}$ makes with the $z$--axis, sampled using a Markov Chain Monte Carlo (MCMC) method as well as the SEM scheme. Further, we can see that while the mean of this distribution is fairly close to $\pi/2$ (for which $\hat{\V{p}}$ is parallel with the walls) the variance is fairly wide. This shows that the cross-point frequently points up or down, which accounts for the bimodality of the cross--point height distribution. The inset in figure \ref{fig:boom}(b) demonstrates that the SEM scheme is first order weakly accurate in both position and orientation statistics as shown in Appendix \ref{appendix}.


\section{Proof of Weak Accuracy for the SEM Scheme} \label{appendix}

As the SEM scheme is an adaptation of the Traction--corrected Euler--Maruyama scheme developed in \cite{BrownianMultiblobSuspensions}, much of proof of first order accuracy from Appendix B2 in \cite{BrownianMultiblobSuspensions} remains valid for the SEM scheme. We may identify $\V{D}^{\K^T}$ from section \ref{tint} of this work with $\V{D}^{F}$ of \cite{BrownianMultiblobSuspensions}. With this identification, all that remains to show is that $\av{\V{D}^{\S}+\V{D}^{\J}-\V{D}^{\K}}$ is equal to $\av{\V{D}^{S}}$ from \cite{BrownianMultiblobSuspensions} and the proof of weak accuracy from Appendix B2 will suffice for the SEM scheme. In the interest of brevity, we will reuse the notation and results already established in Appendix B of \cite{BrownianMultiblobSuspensions}. 

From \eqref{allDemBoysa}--\eqref{allDemBoysb} and \eqref{allDemBoys2a}--\eqref{allDemBoys2b}, we may write
\begin{align}
\av{\V{D}^{\S}_{r}+\V{D}^{\J}_{r}-\V{D}^{\K}_{r}} &= \frac{1}{\delta} \av{\J_{ra} \L^{-1}_{ab} \left[\S_{bs} \left(\V{Q}^{+}\right) - \S_{bs} \left(\V{Q}^{-}\right) \right] {\V{\lambda}}^{RFD}_{s}} \\ &+ \frac{1}{\delta} \av{\left[ \J_{ra} \left(\V{Q}^{+}\right) - \J_{ra} \left(\V{Q}^{-}\right) \right] \V{v}^{RFD}_{a}} \nonumber
\\&-\frac{1}{\delta}\av{\left[ \K_{rl} \left(\V{Q}^{+}\right)  - \K_{rl} \left(\V{Q}^{-}\right) \right] \V{U}^{RFD}_{l}}. \nonumber
\end{align}
Expanding this using \eqref{ulvRFD}, we have in expectation
\begin{align}
\av{\V{D}^{\S}_{r}+\V{D}^{\J}_{r}-\V{D}^{\K}_{r}} &= \frac{1}{\delta} \av{\J_{ra} \L^{-1}_{ab} \left[\S_{bs} \left(\V{Q}^{+}\right) - \S_{bs} \left(\V{Q}^{-}\right) \right] \Mob^{sq}  \K_{ql} \N_{lj} \V{W}_{j}^{\text{FT}} } \\ 
&+ \frac{1}{\delta} \av{\left[ \J_{ra} \left(\V{Q}^{+}\right) - \J_{ra} \left(\V{Q}^{-}\right) \right] \L^{-1}_{ab} \S_{bs} \Mob^{sq}  \K_{ql} \N_{lj} \V{W}_{j}^{\text{FT}}}  \nonumber
\\&-\frac{1}{\delta}\av{\left[ \K_{rl} \left(\V{Q}^{+}\right)  - \K_{rl} \left(\V{Q}^{-}\right) \right] \N_{lj} \V{W}_{j}^{\text{FT}} }  \nonumber \\
&=\left[ \J_{ra} \L^{-1}_{ab} \left(\partial_{k} \S_{bs} \right) +  \left( \partial_{k} \J_{ra} \right) \L^{-1}_{ab} \S_{bs} \right]  \Mob^{sq}  \K_{ql} \N_{lj}\av{\Delta \V{Q}_{k} \V{W}_{j}^{\text{FT}} } \\
&\hspace{2cm}-\left(\partial_{k} \K_{rl} \right) \N_{lj} \av{\Delta \V{Q}_{k} \V{W}_{j}^{\text{FT}} } + \V{O}\left( \delta^2 \right)  \nonumber \\
&=\left[\partial_{k} \left( \J_{ra} \L^{-1}_{ab} \S_{bs} \right)  \Mob^{sq}  \K_{ql} -\left(\partial_{k} \K_{rl} \right) \right] \N_{lj} \av{\Delta \V{Q}_{k} \V{W}_{j}^{\text{FT}} } + \V{O}\left( \delta^2 \right)  \\
&= \left[\left( \partial_{k} \Mob_{rs} \right)  \Mob^{sq}  \K_{ql} -\left(\partial_{k} \K_{rl} \right) \right] \N_{lj} \av{\Delta \V{Q}_{k} \V{W}_{j}^{\text{FT}} } + \V{O}\left( \delta^2 \right)  \\
&=\av{\V{D}^{s}}.
\end{align}
Here we used the definition of $\Mob$ from equation \eqref{MobDefn} in the second to last equality. The last equality simply identified the result with that obtained in equation (B14) of \cite{BrownianMultiblobSuspensions}.
\bibliographystyle{unsrt}
\bibliography{./References,./Sprinkle_Ref}

\end{document}